\documentclass[superscriptaddress,groupedaddress,nofootnoteinbib,11pt]{article}  

\usepackage{graphicx}  
\usepackage{dcolumn}   
\usepackage{bm}        
\usepackage{slashed}        
\usepackage{float}
\usepackage{jcappub}
\usepackage{color}

\usepackage{graphicx}
\usepackage{dcolumn}
\usepackage{bm}
\usepackage{amssymb}
\usepackage{amsmath}
\usepackage{sectsty}
\usepackage{colortbl}
\usepackage{latexsym}
\usepackage{float}
\usepackage{ifthen}
\usepackage{enumerate}
\usepackage{url}
\usepackage[force]{feynmp-auto}
\usepackage{jcappub}
    \usepackage{picinpar}
    \usepackage{colortbl}
\usepackage{multirow}
	\usepackage{float}
	      \usepackage{setspace}
\usepackage{array}
\usepackage{bm}
\usepackage{amsopn}

\usepackage{caption}
\usepackage{subcaption}

\usepackage{booktabs}
\usepackage[table]{xcolor}

\def\be{\begin{equation}}
\def\ee{\end{equation}}
\def\bea{\begin{eqnarray}}
\def\eea{\end{eqnarray}}
\def\beq{\begin{eqnarray}}
\def\eeq{\end{eqnarray}}

\def\L*{{\cal L}_*}
\def\L{\mathcal{L}}
\def\({\left(}
\def\){\right)}

\def\<{\langle}
\def\>{\rangle}

 \def\neq {\not\equiv}

\def\cs2{c_{s}^{2}}

 \def\be   {\begin{equation}}   \def\ee   {\end{equation}}
 \def\ba  {\begin{eqnarray}}   \def\ea  {\end{eqnarray}}
 \def\bean {\begin{eqnarray*}}  \def\eean {\end{eqnarray*}}

\definecolor{RoyalBlue}{rgb}{0.25,.41,.88}
\definecolor{RedWine}{rgb}{0.743,0,0}

\begin{document}

\title{Reheating predictions in single field inflation}

\author[a]{Jessica L. Cook,}
\author[a]{Emanuela Dimastrogiovanni,}
\author[a]{Damien A. Easson,}
\author[a,b]{Lawrence M. Krauss}
\affiliation[a]{Department of Physics and School of Earth and Space Exploration \\ Arizona State University, Tempe, AZ 85827-1404}
\affiliation[b]{Research School of Astronomy and Astrophysics, Mt. Stromlo Observatory, \\ Australian National University, Canberra, Australia 2611}

\emailAdd{jlcook14@asu.edu}
\emailAdd{emad@asu.edu}
\emailAdd{easson@asu.edu}
\emailAdd{krauss@asu.edu}
\date{\today}
\abstract{

  Reheating is a transition era after the end of inflation, during which the inflaton is converted into the particles that populate the Universe at later times. No direct cosmological observables are normally traceable to this period of  reheating. Indirect bounds can however be derived. One possibility is to consider cosmological evolution for observable CMB scales from the time of Hubble crossing to the present time. Depending upon the model, the duration and final temperature after reheating, as well as its equation of state, may be directly  linked to inflationary observables. For single-field inflationary models, if we approximate reheating by a constant equation of state, one can derive relations between the reheating duration (or final temperature), its equation of state parameter, and the scalar power spectrum amplitude and spectral index. While this is a simple approximation, by restricting the equation of state to lie within a broad physically allowed range, one can in turn bracket an allowed range of $n_s$ and $r$ for these models. The added constraints can help break degeneracies between inflation models that otherwise overlap in their predictions for $n_s$ and $r$.

 }
\maketitle

\section{Introduction}

The inflationary paradigm \cite{Starobinsky:1980te,Guth:1980zm,Sato:1980yn,Linde:1981mu,Albrecht:1982wi,Linde:1983gd,Lyth:1998xn,Riotto:2002yw,Kinney:2003xf,Baumann:2009ds} offers, in its numerous constructions (see e.g. \cite{Martin:2014vha}), a testable \cite{Planck:2015xua,Ade:2015oja} description for the physics of the very early Universe. Inflation addresses several open problems in cosmology, chief among them the question of the origin of cosmological structures. In its simplest realization, the Universe is dominated by the potential energy of a light scalar field, the inflaton, that drives the expansion. In this picture, quantum fluctuations of the scalar field during inflation are precisely the primary source of cosmological perturbations \cite{Mukhanov:1981xt,Starobinsky:1982ee,Guth:1982ec,Bardeen:1983qw,Abbott:1984fp,Abbott:1984qq}. The statistical properties of the Cosmic Microwave Background (CMB) fluctuations and of the Large Scale Structures (LSS) may therefore contain information about the physics of inflation. In addition to scalar density perturbations, inflation generically produces tensor perturbations, resulting in a spectrum of primordial gravitational waves which, via their impacts on the CMB and other astronomical sources, reveal information about inflation {\cite{Starobinsky:1979ty,Rubakov:1982df,Krauss:1992ke,Krauss:2013pha}. \\

The transition from inflation to later stages of the evolution of the Universe (radiation and matter dominance) is referred to as \textsl{reheating}. During reheating the inflaton field loses its energy, eventually leading to the production of ordinary matter. Several reheating models have been proposed: the simplest ones, involve the perturbative decay of an oscillating inflaton field at the end of inflation \cite{Abbott:1982hn,Dolgov:1982th,Albrecht:1982mp}, while more intricate scenarios include non-perturbative processes such as (broad) parametric resonance decay \cite{Kofman:1994rk,Traschen:1990sw,Kofman:1997yn}, tachyonic instability \cite{Greene:1997ge,Shuhmaher:2005mf,Dufaux:2006ee,Abolhasani:2009nb,Felder:2000hj,Felder:2001kt}, and instant preheating \cite{Felder:1998vq}\footnote{See also \cite{Boyanovsky:1996sv,Bassett:2005xm,Allahverdi:2010xz,Amin:2014eta} for reviews {and, e.g., \cite{Drewes:2013iaa,Drewes:2014pfa} for more studies on reheating}.}. The word \textsl{preheating} indicates the initial stage of reheating, especially in the context where decay happens exponentially, generating high occupation numbers in select frequency bands. Immediately after preheating the frequency bands that underwent parametric resonance will have extremely high occupation numbers while the rest of the space will be basically un-populated, a highly non-thermal state. Over time, scattering events will spread out the distribution, eventually leading to a blackbody spectrum characterized by a final temperature $T_{re}$, which normally corresponds to the temperature at the beginning of the radiation-dominated era. \\

\indent For some inflationary scenarios and for given interactions between the inflaton field and other matter fields, numerical studies were performed to derive an effective equation of state (eos). The eos is parametrized by a function $w_{re}(t)$ for the Universe during the various stages of reheating. As inflation ends, the eos parameter is equal to $-1/3$. Assuming a massive inflaton, very quickly the eos climbs to 0, the eos of a massive harmonic oscillator oscillating between potential dominance (eos of $-1$) and kinetic dominance (eos of $1$). During this initial phase of reheating, the frequency of oscillations, characterized by the inflaton mass $m$, will be larger than the expansion rate. It is therefore correct to approximate the eos of the inflaton as a constant of 0. This is the equation of state of the Universe at the beginning of reheating when the Universe is still dominated by the inflaton field. As the inflaton decays and the decay products compose an increasing percentage of the energy density of the Universe, the eos will increase from 0 to $1/3$ at the start of radiation dominance. 
In \cite{Podolsky:2005bw} it was shown that for a simple chaotic inflation model and for a quartic $g^{2}\phi^{2}\chi^{2}$ interaction ($\phi$ being the inflation and $\chi$ its decay product), the equation of state right after inflation, characterized by $w_{re}=0$, sharply, within a couple efolds, changes to $w_{re}\sim 0.2-0.3$ already during preheating, long before the system reaches thermal equilibrium\footnote{A physical system reaching an effective (macroscopic) state characterized by nearly constant ratio of pressure over energy density while it is, microscopically, still out-of-equilibrium (``pre-thermalization'') had been previously investigated in Minkowski spacetime in \cite{Berges:2004ce}.}. The duration of preheating can therefore generally be regarded as ``instantaneous'' in comparison with the remaining stages of reheating. In cases like the ones described in \cite{Podolsky:2005bw} (see also \cite{Kofman:1994rk}), $w_{re}$ may therefore be rightfully treated as a constant throughout the entire reheating era. \\
\indent Aside from its thermalization temperature, $T_{re}$, and effective equation of state, $w_{re}$, reheating is also characterized by its duration, which one may quantify in terms of e-foldings $N_{re}\equiv \ln (a_{re}/a_{end})$, occurring between the time inflation ends, $t_{end}$, and the beginning of the radiation-dominated era, $t_{re}$. \\

The reheating era is a difficult one to constrain observationally: except for some non-conventional scenarios (e.g. \cite{Taruya:1997iv,Bassett:1999cg,Finelli:2000ya,Tsujikawa:2002nf,Chambers:2007se,Bond:2009xx,Bassett:1998wg,Bassett:1999mt,Bassett:1999ta,Bethke:2013aba,Easther:2013nga,Moghaddam:2014ksa}).  In the absence of topological defects like monopoles or strings, the fluctuations produced during reheating remain sub-horizon and cannot leave an observable imprint at the level of the CMB or LSS. A lower bound is placed on the reheating temperature by primordial nucleosynthesis (BBN) $T_{BBN}\sim 10^{-2}GeV$ \cite{Steigman:2007xt}\footnote{{Smaller values may be assigned to the lower bound of the reheating temperature in models such as \cite{Kawasaki:1999na}}.}; the scale of inflation is merely bounded from above (the CMB B-modes recently measured by BICEP2 \cite{Ade:2014xna,Ade:2014gua} do not yet, unfortunately, point to an inflationary signal) and can be as large as $\sim 10^{16}GeV$, leaving for $T_{re}$ an allowed range of many orders of magnitude . Aside for the production of metric fluctuations in the aforementioned scenarios, a variety of signatures (or lack thereof) relative to the production of primordial black holes \cite{GarciaBellido:1996qt,Carr:2009jm,Torres-Lomas:2014bua}, magnetic field \cite{Calzetta:2001cf,DiazGil:2007dy,DiazGil:2008tf}, unwanted relics \cite{Giudice:1999yt,Giudice:2001ep} and also to mechanisms such as baryo-and leptogenesis \cite{Giudice:1999fb,Krauss:1999ng,GarciaBellido:1999sv,Davidson:2000dw,Copeland:2001qw} (and more, see \cite{Allahverdi:2010xz} for an overview and for a full list of related references), may be traced back to specific preheating/reheating models. \\
 
Another possibility for extracting information about reheating is to consider the expansion history of the Universe between the time the observable CMB scales crossed outside the Hubble radius during inflation and the time they later re-entered, in such a way as to define a relation between inflationary and reheating parameters \cite{Liddle:2003as}
\begin{equation}\label{prima}
\ln\left[\frac{k}{a_{0}H_{0}}\right]=-N_{k}-N_{re}-N_{RD}+\ln\left[\frac{a_{eq}H_{eq}}{a_{0}H_{0}}\right]+\ln\left[\frac{H_{k}}{H_{eq}}\right].
\end{equation}
In this equation, $k$ can be chosen as the pivot scale for a specific experiment, $N_{k}$ is the number of e-foldings between the exit time of the modes at this pivot during inflation and the end of inflation, $N_{re}$ and $N_{RD}$ respectively indicated the e-folds between the end of inflation and the end of reheating and between the end of reheating and the end of the radiation-dominated era. From (\ref{prima}) one realizes that from the CMB constraints on the primordial power spectrum (which would correspond to a prediction for $N_{k}$), for a given inflationary model one would be able to infer the sum of $N_{RD}$ and $N_{re}$. To solve for $N_{re}$ and $N_{RD}$ individually one needs more information. For reheating models that can be parametrized by a constant effective pressure to energy ratio $w_{re}$,  one can relate the density at the end of inflation to the density at the end of reheating, and then assuming conservation of entropy after reheating, to the temperature today. This way one obtains another equation with the same two unknowns $N_{re}$ and $N_{RD}$ that can be used to solve for each individually, or to rework the equations to trade the quantity $N_{re}$ for $T_{re}$, the temperature at the end of reheating. All of this is particularly straightforward for single-field models of inflation that are entirely defined by the form of their potential. In summary, for a given inflationary model and for given equations of state during reheating lying within a reasonable physically plausible range, one may use the CMB data to place constraints on the reheating temperature and its duration. These techniques have been successfully employed in several studies \cite{Martin:2006rs,Lorenz:2007ze,Martin:2010kz,Adshead:2010mc,Mielczarek:2010ag,Easther:2011yq,Dai:2014jja,Martin:2014nya}.  \\

{In the same spirit as \cite{Martin:2006rs,Lorenz:2007ze,Martin:2010kz,Adshead:2010mc,Mielczarek:2010ag,Easther:2011yq,Dai:2014jja,Martin:2014nya}, and using similar techniques as in \cite{Dai:2014jja}} (where the attention was directed specifically to inflation with power-law potentials, $V(\phi)\sim \phi^{\alpha}$), we consider the constraints imposed by reheating on popular single field inflationary scenarios. We derive predictions for the length of the reheating era, and the temperature at the end of reheating for each model, assuming a constant equation of state during reheating. Accounting for the lower bounds on $T_{re}$ imposed by BBN and considering a physically plausible range of values for $w_{re}$ (likely the average value will fall between 0 and $\frac{1}{3}$) we use the relations between reheating and inflationary parameters and the constraints on the primordial power spectrum amplitude and tilt from Planck \cite{Planck:2015xua,Ade:2015oja} to provide new constraints on the parameter space in given inflationary models. This is a useful and relatively new tool for constraining and differentiating between inflation models.  Models might overlap in predictions for $n_s$ and $r$, but not for the same $w_{re}$. As the constraints on $n_s$ gets tighter, this will translate into an increasingly narrow allowed range for $w_{re}$ for a given inflation model, and so this technique of constraining models with reheating will be increasingly efficient in ruling out some models in favor of others. \\

This work is organized as follows: in Sec.~\ref{sec2} we detail the derivation of the reheating duration and of the temperature at the end of reheating as a function of the spectral index, for canonical single-field inflationary models and for reheating scenarios that can be described in terms of a constant effective equation of state; in Sec.~\ref{sec3} we review the analysis of \cite{Dai:2014jja} for a power law potential and we discuss the constraints from reheating on the inflationary parameters; in Secs.~\ref{sec4} through \ref{sec7} we compute the relations between inflationary and reheating parameters in the Starobinsky, Higgs, natural and hilltop inflation models and we discuss the bounds placed on some of these models by reheating; in Sec.~\ref{sec8} we present our conclusions.

\section{Calculating $N_{re}$ and $T_{re}$}
\label{sec2}

A reheating model (or class of models) may be characterized by a thermalization temperature $T_{re}$, a duration, $N_{re}$ (here defined in terms of the number of e-folds counted from the end of inflation), and an equation of state with an effective pressure-to-energy-density ratio, $w_{re}$. The latter should have values larger than $-1/3$ for inflation to come to an end, and is assumed to be smaller than $1$ in order not to violate causality. A variety of reheating scenarios allow for an equation of state that is nearly constant in time. For the purposes of this work we will thus approximate $w_{re}$ as a constant in all our calculations; in our plots for $N_{re}$ and $T_{re}$, we assign to $w_{re}$ sample values ranging in the interval $[-1/3,1]$. We define $N_{re}$ as the time frame from the end of inflation until the equation of state makes a step function transition from the value $w_{re}$ it had during reheating to $w =1/3$, which we define as the start of radiation dominance. $T_{re}$ is the temperature when this transition occurs. From this definition, $N_{re}$ and $T_{re}$ are not well defined if the equation of state during reheating is also equal to $1/3$ (we will discuss this case more later). Also, we assume a standard expansion history after reheating, with a radiation-dominated (RD) era followed by a matter-dominated (MD) one. 

We derive, following \cite{Martin:2006rs,Lorenz:2007ze,Martin:2010kz,Adshead:2010mc,Mielczarek:2010ag,Easther:2011yq,Dai:2014jja,Martin:2014nya}, an expression for the reheating parameters ($N_{re}$, $T_{re}$ and $w_{re}$) in terms of a set of physical quantities that are specific to inflation and to the cosmological epochs subsequent to reheating. Considering the evolution of the Universe between the Hubble-exit time during inflation (henceforth indicated by $t_{k}$) for observable scales and the time of observation of the same scales ($t_{0}$), one can write matching conditions for the total energy density as well as for the scale factor, $a(t)$, during the intermediate eras. Fig.~(\ref{fig:gg}) summarizes the evolution of the comoving horizon distance throughout this length of time, marked by the transitions between consecutive epochs at $t_{end}$, the end of inflation, $t_{re}$, the end of reheating/beginning of RD era, and $t_{eq}$, the beginning of the MD era.

In the figure we equate the size the comoving horizon far back into inflation, corresponding to modes $l=2$, to the size of the horizon today. In order to solve the horizon problem, the span of comoving scales that leave the horizon from $l=2$ to the end of inflation must equal the span of comoving scales that reenter the horizon after inflation till today. Note the factor by which the comoving horizon shrinks between scales $l=2$ and the end of inflation (the length of the first line in the figures) is not known. The slope of that line is set by the fact that the equation of state is $\approx -1$ during inflation.  Depending on the model, that line could be longer or shorter. While there is a minimum length in order to solve the horizon problem while having Inflation occur before BBN, there is no upper bound. The value of $w_{re}$ will set the slope of the second line, the rate by which modes reenter the horizon during reheating. In the figure we display the two extreme cases of $w_{re} =1$ and $w_{re} = - 1/3$. One can see from comparing the two plots, the smaller $w_{re}$ is during reheating, the less efficiently modes re-enter the horizon, and the more efolds will be necessary in the post-inflation period. 

We consider single-field inflationary models with background field equations, $\ddot{\phi}+3H\dot{\phi}+V^{'}=0$ and $3H^{2}M_{P}^{2}\simeq V(\phi)$. We also assume that both $\epsilon$ and $\eta$ remain smaller than 1 throughout the inflationary regime.\\

\begin{figure}
\centering
\begin{subfigure}{.5\textwidth}
  \centering
  \includegraphics[width=.95\linewidth]{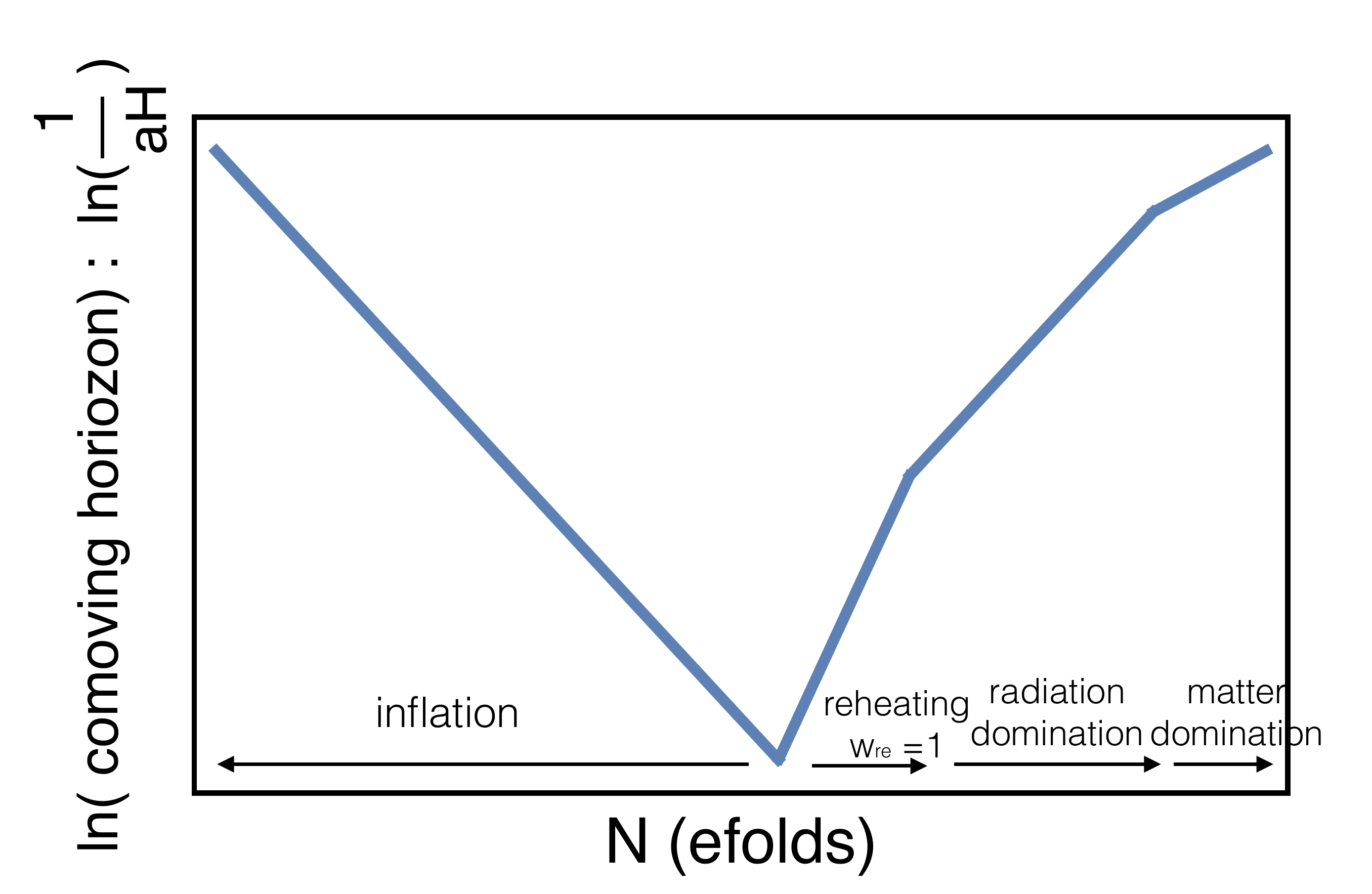}
\end{subfigure}%
\begin{subfigure}{.5\textwidth}
  \centering
  \includegraphics[width=.9\linewidth]{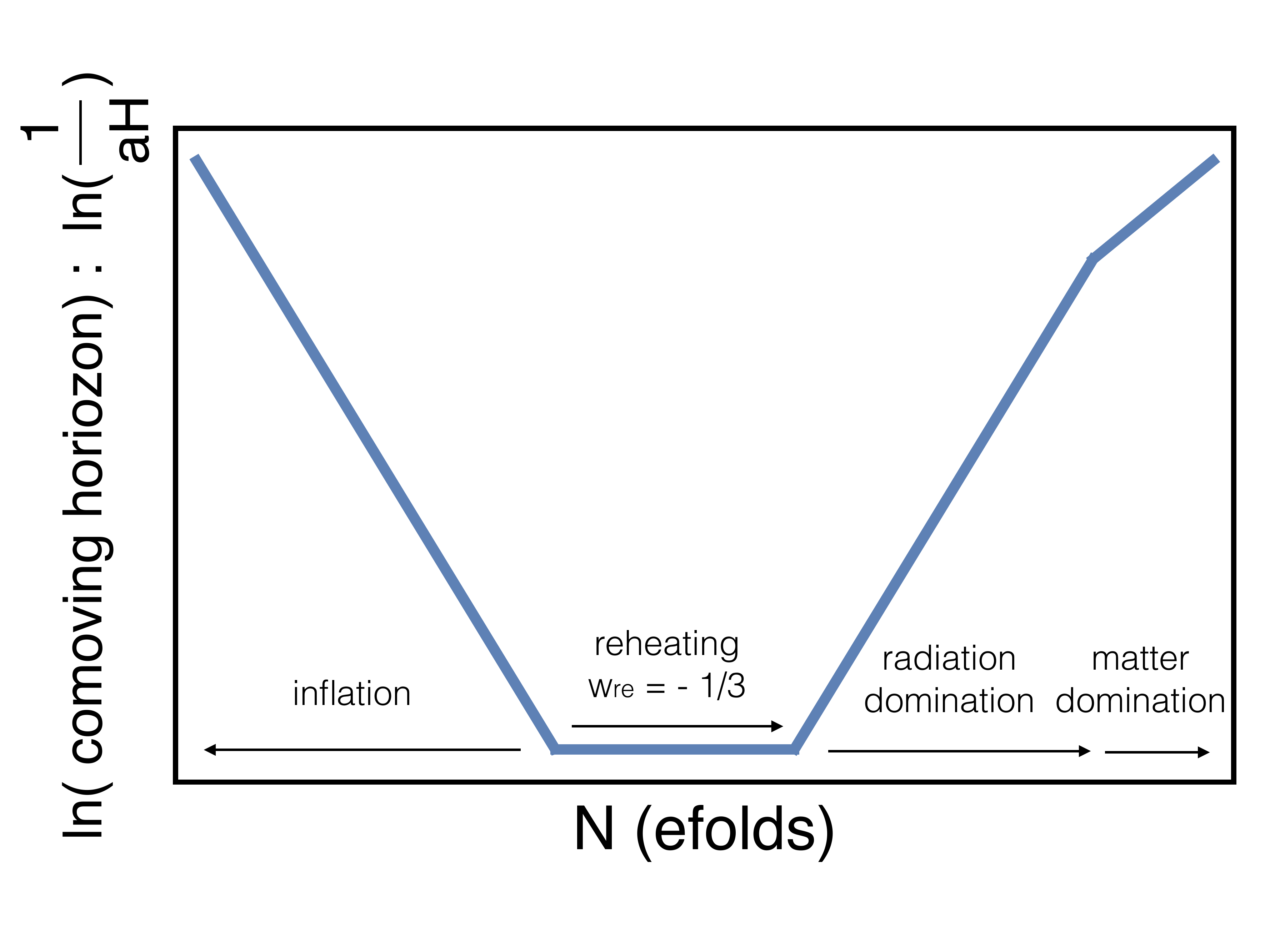}
\end{subfigure}
\caption{Each figure shows the evolution of the comoving horizon distance over time. Each figure shows the extreme cases for $w_{re}$: the first figure for $w_{re} = 1$ and the second for $w_{re} = - \frac{1}{3}$.}
\label{fig:gg}
\end{figure}

\noindent If one assumes a constant equation of state, the change in the scale factor during reheating is easily related to the change in the energy density.  Using $\rho \propto a^{-3(1+w)}$, the reheating epoch is described by
\begin{align}
\frac{\rho_{end}}{\rho_{re}} = \left(\frac{a_{end}}{a_{re}} \right)^{-3(1+w_{re})},
\end{align}
where the subscript $end$ refers to the end of inflation (the start of reheating), and $re$ refers to the end of reheating. Writing this in terms of e-foldings
\begin{align}\label{eqq2}
N_{re} = \frac{1}{3(1+w_{re})} \ln \left(\frac{\rho_{end}}{\rho_{re}} \right)= \frac{1}{3(1+w_{re})} \ln \left(\frac{3}{2}\frac{V_{end}}{\rho_{re}} \right),
\end{align}
where the last step of (\ref{eqq2}) is obtained by replacing $\rho_{end} = (3/2) V_{end}$, derived by setting $w = - 1/3$ at the end of inflation.\\
The temperature is related to the density by
\begin{align}\label{eqq3}
\rho_{re} = \frac{\pi^2}{30} g_{re} T_{re}^4,
\end{align}
where $g_{re}$ is the number of relativistic species at the end of reheating. Combining Eqs.~(\ref{eqq2}) and (\ref{eqq3}) one finds
\begin{align}\label{eq2}
N_{re} = \frac{1}{3(1+w)} \ln \left(\frac{30 \cdot \frac{3}{2}  V_{end}}{\pi^2 g_{re} T_{re}^4 } \right).
\end{align}
Making the standard assumption that entropy is conserved between the end of reheating and today, one can relate the reheating temperature to the temperature today by taking into account the changing number of helicity states in the radiation gas as a function of temperature,
\begin{align}\label{eq6}
T_{re}= T_0 \left(\frac{a_0}{a_{re}} \right) \left(\frac{43}{11 g_{re}} \right)^{\frac{1}{3}}=T_0 \left(\frac{a_0}{a_{eq}} \right) e^{N_{RD}} \left(\frac{43}{11 g_{re}} \right)^{\frac{1}{3}},
\end{align}
where $N_{RD}$ is the length in e-folds of radiation dominance, $e^{-N_{RD}}\equiv a_{re}/a_{eq}$. The ratio $a_{0}/a_{eq}$ can be rewritten as
\begin{align}\label{eq3}
\frac{a_0}{a_{eq}} = \frac{a_0 H_{k}}{k} e^{-N_{k}} e^{- N_{re}} e^{- N_{RD}}\, ,
\end{align}
where one uses the relation $k_{}=a_{k} H_{k}$ for the time at which the pivot scale k \footnote{Note in the following when we repeatedly refer to the pivot scale, we will use throughout Planck's pivot scale of $0.05 Mpc^{-1}$.} crosses outside the Hubble radius and $N_{k}$ is defined as the number of e-foldings between the latter and the time inflation ends. Inserting (\ref{eq3}) into (\ref{eq6}) one finds 
\begin{align}\label{eq8}
T_{re} =  \left(\frac{43}{11 g_{re}} \right)^{\frac{1}{3}}  \left(\frac{a_0 T_0}{k_{}} \right) H_{k} e^{-N_{k}} e^{- N_{re}}.
\end{align}
Notice that larger values of $N_{re}$ corresponds to smaller $T_{re}$ and vice versa. In other words, as expected, the quicker and more efficiently reheating takes place, the larger the temperature. Plugging (\ref{eq8}) into Eq.~(\ref{eq2}) 
\begin{align}\label{eq4}
 N_{re} = \frac{4}{3 (1+ w_{re})}   \left[ \frac{1}{4}  \ln \left(\frac{3^2 \cdot 5}{\pi^2 g_{re}} \right) + \ln \left(\frac{V_{end}^{\frac{1}{4}}}{H_{k}} \right) + \frac{1}{3} \ln \left(\frac{11 g_{re}}{43} \right) + \ln \left(\frac{k_{}}{a_0 T_0} \right) + N_{k} + N_{re}   \right].
\end{align}
One can first solve for $N_{re}$ assuming $w_{re} \neq \frac{1}{3}$
\begin{align}\label{align}
 N_{re}= \frac{4}{ (1-3w_{re} )}   \left[- \frac{1}{4}  \ln \left(\frac{3^2 \cdot 5}{\pi^2 g_{re}} \right) - \frac{1}{3} \ln \left(\frac{11 g_{re}}{43} \right) - \ln \left(\frac{k_{}}{a_0 T_0} \right)  - \ln \left(\frac{ V_{end}^{\frac{1}{4}}}{ H_{k} } \right)  - N_{k}   \right] \, .
\end{align}
Notice that the values of the last two terms in Eq.~(\ref{align}) depend on the specific inflationary model. Assuming $g_{re} \approx 100$ and using Planck's pivot of $0.05 Mpc^{-1}$ \footnote{The convention in the Planck analysis defines the pivot scale such that the comoving momentum $k$ becomes horizon sized when $k a_0 = a H$, where we have been using $k = a H$, so using our conventions $\frac{k_{}}{a_0} =  0.05 Mpc^{-1}$.}, one obtains a simplified expression for $N_{re}$, before specifying a particular inflationary model:
\begin{align}\label{eq12}
 N_{re}= \frac{4}{ (1-3w_{re} )}   \left[61.6  - \ln \left(\frac{ V_{end}^{\frac{1}{4}}}{ H_{k} } \right)  - N_{k}   \right].
\end{align}
One can then use Eq.~(\ref{eq8}) to obtain 
\begin{align}\label{mp1}
T_{re}= \left[ \left(\frac{43}{11 g_{re}} \right)^{\frac{1}{3}}    \frac{a_0 T_0}{k_{}} H_{k} e^{- N_{k}} \left[\frac{3^2 \cdot 5 V_{end}}{\pi^2 g_{re}} \right]^{- \frac{1}{3(1 + w_{re})}}  \right]^{\frac{3(1+ w_{re})}{3 w_{re} -1}}.
\end{align}

\subsection{Special case $w_{re}=\frac{1}{3}$}

The final result for $N_{re}$ in Eq.~(\ref{eq12}) only applies for $w_{re} \neq 1/3$. Going back to Eq.~(\ref{eq4}), notice that if $w_{re} = \frac{1}{3}$, $N_{re}$ cancels from both sides of the equation, and one is left with
\begin{align}
 0 =  \frac{1}{4}  \ln \left(\frac{30}{\pi^2 g_{re}} \right) + \frac{1}{4} \ln \left(\frac{3}{2} \right)  +  \ln \left(\frac{ V_{end}^{\frac{1}{4}}}{H_{k}} \right) + \frac{1}{3} \ln \left(\frac{11 g_{re}}{43} \right) + \ln \left(\frac{k_{}}{a_0 T_0} \right)       +  N_{k} \,.
\end{align}
Assuming $g_{re} = 100$, and Planck's pivot scale, this simplifies to:
\begin{align}\label{eq5}
 61.6 =   \ln \left(\frac{ V_{end}^{\frac{1}{4}}}{H_{k}} \right) + N_{k} \,.
\end{align}
For $w=1/3$, it is not possible to derive a prediction for $N_{re}$ or $T_{re}$ but instead, for a particular inflation model, one finds a prediction for $n_s$. Note the ambiguity in $N_{re}$ and $T_{re}$ is due to the fact that we are defining the start of radiation dominance as the moment $w_{re}$ reaches $1/3$. If $w_{re}$ is already equal to $1/3$ during reheating, then there is ambiguity in when to differentiate between the two regimes.


\subsection{Model dependent part}

In order to solve for $N_{re}$  in Eq.~(\ref{eq12}) (or to solve for $n_s$ in Eq.~(\ref{eq5}) if $w_{re} = 1/3$) for a particular model, one needs to compute $N_{k}$, $H_{k}$, and $V_{end}$. $N_{k}$ can be calculated starting from the definition of e-foldings:
\begin{align}\label{eqq1}
\Delta N = \int H dt\,.
\end{align}
Recasting the r.h.s. of (\ref{eqq1}) as an integral over $\phi$ and using the background equation of motion for the inflaton, $3 H \dot{\phi} + V' \simeq 0$, and the Friedmann equation, $H^2 \simeq V/(3 M_P^2)$, one finds
\begin{align}\label{eq13}
N_k \simeq \frac{1}{M_P^2} \int_{\phi_{end}}^{\phi_k} \frac{V}{V'}\, d \phi\,.
\end{align}
Next, $H_{k}$ can be written as a function of $n_s$. Using the definition of the tensor-to-scalar ratio $r = P_h/P_{\zeta}$ (where $P_h =(2 H^2)/(\pi^2 M_P^2)$ and $P_{\zeta} = A_s$ at the pivot scale) 
\begin{align}
r_{k} = \frac{2 H_k^2}{ \pi^2 M_P^2 A_s}.
\end{align}
Then using $r =16 \epsilon$ this gives
\begin{align}\label{eq14}
H_{k} \simeq \pi M_P \sqrt{8 A_s \epsilon_{k}}.
\end{align}
Once the form of $V(\phi)$ is specified for a given model, one can express $V_{end}$ as a function of model parameters calculated at the pivot scale. The explicit form of $V_{end}$ along with (\ref{eq13}) and (\ref{eq14}) can be plugged into Eqs.~(\ref{eq12}) and (\ref{mp1}) to derive $N_{re}$ and $T_{re}$ as a function of inflationary model parameters (or into Eq.~(\ref{eq5}) in the case $w_{re} = 1/3$).


\section{Polynomial potentials}
\label{sec3}

{Consider a polynomial type potential
\begin{align}\label{eqqq1}
V = \frac{1}{2} m^{4 - \alpha} \phi^{\alpha}.
\end{align}
This was considered in the context of reheating in \cite{Martin:2006rs,Martin:2010kz,Dai:2014jja,Martin:2014nya,Martin:2014vha}}. We quickly review this specific application. At the end of this section, we discuss with some quantitative examples how closely the constraints from inflation compare to the ones from reheating.\\ 
The first step is to calculate the model dependent parameters in Eq.~(\ref{eq12}), i.e. $N_k$, $H_k$, and $V_{end}$. The number of e-folds between the time the pivot scale exited the Hubble radius and the end of inflation can be derived using Eq.~(\ref{eq13})
\begin{align}
N_k = \frac{1}{2 \alpha M_P^2} \left(\phi_k^2 - \phi_{end}^2 \right) .
\end{align}
The potential in these polynomial models is generally steep enough so that $\phi_k \gg \phi_{end}$ and it is appropriate to approximate
\begin{align}\label{eqq33}
N_k \approx \frac{1}{2 \alpha M_P^2} \phi_k^2\, .
\end{align}
We now require $N_k$ as a function of $n_s$. From the expression of the spectral index as a function of the slow-roll parameters, $n_s= 1 - 6 \epsilon + 2 \eta$ (where $\epsilon = (M_P^2/2)(V'/V)^2$ and $\eta = M_P^2 V''/V$), and using (\ref{eqq33}) to rewrite $\epsilon$ and $\eta$ as functions of $N_k$, one finds 
\begin{align}
N_k = \frac{\alpha + 2}{2 (1 - n_{s})}.
\end{align}
From Eq.~(\ref{eq14}) and using the previous equation, $H_k$ is given by
\begin{align}
H_k =\pi M_{P}\sqrt{\frac{4\pi A_{s}}{\alpha+2}(1-n_{s})}.
\end{align}
Lastly one computes $V_{end}$ in terms of $n_s$ and $A_s$,%
\begin{align}\label{eqqq2}
V_{end}=3 M_P^2 H_k^2  \frac{\phi_{end}^{\alpha}}{\phi_k^{\alpha}}= 6 \pi^2 M_P^4 A_s (1- n_s) \left (\frac{\alpha (1- n_s)}{2 (\alpha + 2)} \right),
\end{align}
where the value of the inflaton field at the end of inflation was computed by solving for $\phi_{end}$ from the condition $\epsilon = 1$.\\

Thus $N_k$, $H_k$, and $V_{end}$ are all expressed as functions only of $\alpha$, $n_s$ and $A_s$ and one may plot $N_{re}$ (and $T_{re}$) as a function of $n_{s}$ for some fixed values of $w_{re}$ and $\alpha$. We use $n_{s}=0.9682\pm0.0062$ and Planck's central value $A_{s}=2.196\times 10^{-9}$ (small variations in $A_{s}$ have negligible effects on reheating predictions). \\


\begin{figure} 
\centering
    \includegraphics[width=16cm]{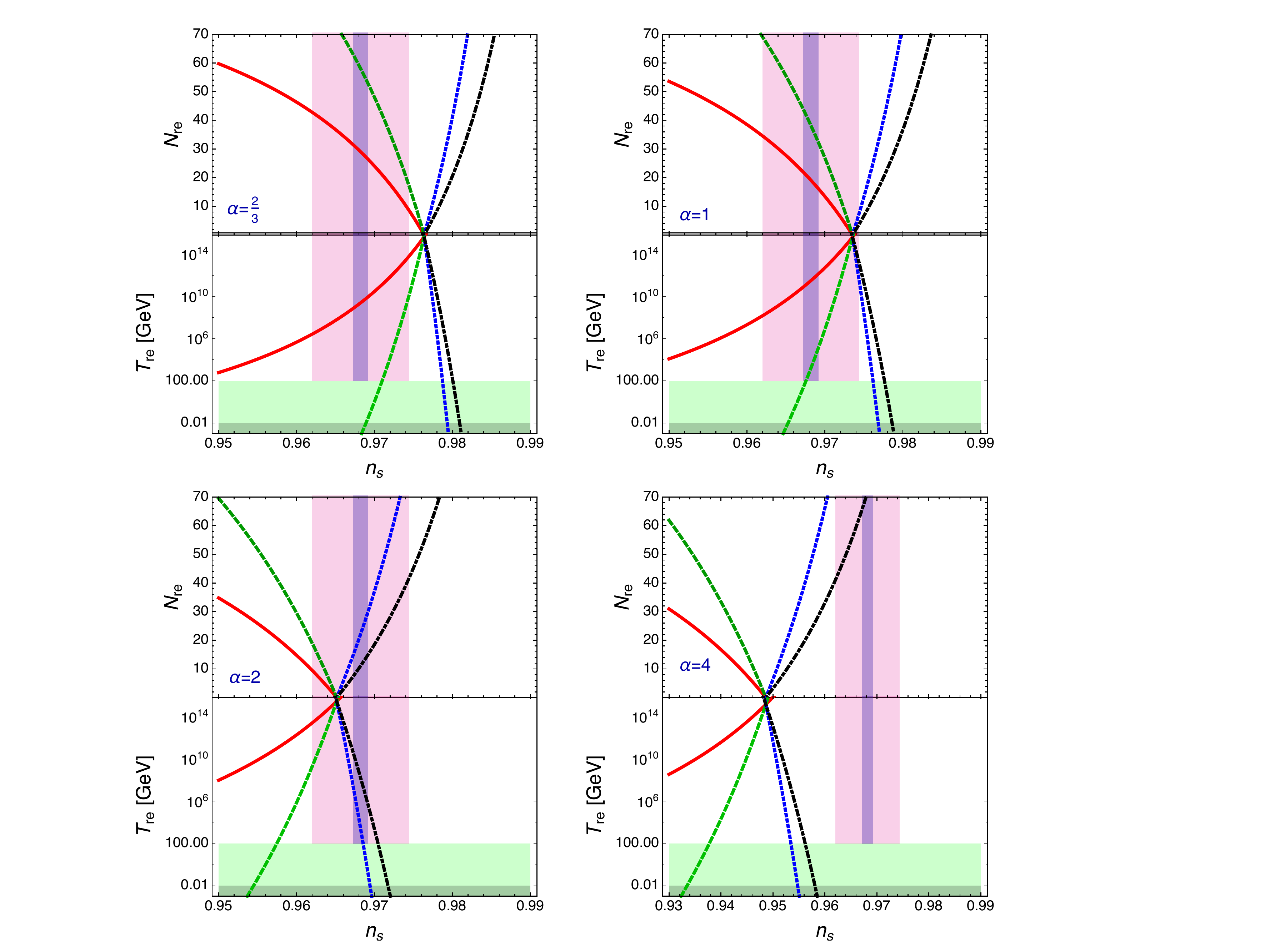}
\caption{Plots of $N_{re}$ and $T_{re}$, the length of reheating and the temperature at the end of reheating respectively, for polynomial potentials with exponent $\alpha$. The solid red line corresponds to $w_{re} = -1/3$, the dashed green line to $w_{re} = 0$, the dotted blue line to $w_{re} = 2/3$, and the dot-dashed black line to $w_{re} =1$. The pink shaded region corresponds to the $1 \sigma$ bounds on $n_s$ from Planck. The purple shaded region corresponds to the $1 \sigma$ bounds of a further CMB experiment with sensitivity $\pm 10^{-3}$ \cite{Amendola:2012ys,Andre:2013afa}, using the same central $n_s$ value as Planck. Temperatures below the dark green shaded region are ruled out by BBN. The light green shaded region is below the electroweak scale, assumed 100 GeV for reference. This region is not disallowed but would be interesting in the context of baryogenesis.}
\label{fig:a}
\end{figure}

We plot in Fig.~\ref{fig:a} $N_{re}$ and $T_{re}$ predictions for $\alpha = 2/3$, 1, 2 and 4. The case $\alpha = 2/3$ is favored by axion-monodromy models, and $\alpha=1$ and $\alpha =2$ give promising predictions when compared with the Planck data. The case $\alpha =4$ is difficult to reconcile with $w_{re}\leq 1$ even considering the 2$\sigma$ bounds on $n_{s}$\footnote{{An exception where $\phi^{4}$ may still be viable is in the context of warm inflation \cite{Bartrum:2013fia,Bastero-Gil:2014oga}.}}. \\
Instantaneous reheating is defined as the limit $N_{re} \rightarrow 0$, visualized in the figure as the point where all the lines converge. Such instantaneous reheating leads to the maximum temperature at the end of reheating, and the equation of state parameter is irrelevant. \\ 
(Thus, while not shown, a $w_{re} = \frac{1}{3}$ solution would correspond to a vertical line passing through the instantaneous reheat point.) 

From Fig.~\ref{fig:a},  $\alpha =2/3$ can be consistent with Planck bounds, but assuming an equation of state $w_{re} \geq 0$, the model would tend to predict smaller reheating temperatures if one considers Planck's $1 \sigma$ bound on $n_{s}$; using Planck's 2$\sigma$ bounds, any reheating temperature up to the maximum instantaneous case is still allowed.

For $\alpha = 1$ and $\alpha = 2$ all the lines in Fig.~\ref{fig:a} are shifted towards the central value of $n_{s}$  when compared to the $\alpha=2/3$ case, thus allowing for a wider range of reheating temperatures as well as values of the equation of state parameter.\\

Consider now the case $w_{re} =1/3$. Solving Eq.~(\ref{eq5}) for the polynomial potential, one obtains
\begin{align}\label{specific1}
61.6= \frac{1}{4} \ln \left(\frac{3 \alpha}{4 \pi^2 A_s (\alpha + 2)}\right) + \frac{\alpha + 2}{2 (1-n_s)}.
\end{align}
Using Planck's central value for $A_s$, Eq.~(\ref{specific1}) gives specific predictions for $n_s$
\begin{equation}
\begin{cases}
n_s = 0.977 \quad\quad  \text{for} \quad\alpha = \frac{2}{3}\,,\\
n_s = 0.974   \quad\quad \text{for} \quad\alpha = 1\,,\\
n_s = 0.965   \quad\quad \text{for} \quad\alpha = 2\,.\\
\end{cases}
\end{equation}
Notice that larger values of $\alpha$ require smaller values of $n_s$. With the $2 \sigma$ bounds on $n_s$ from Planck, $0.956 < n_s < 0.981$,  $w_{re} = \frac{1}{3}$ would be consistent with all three values of $\alpha$.\\

\begin{figure}[H]
\centering
    \includegraphics[width=14cm]{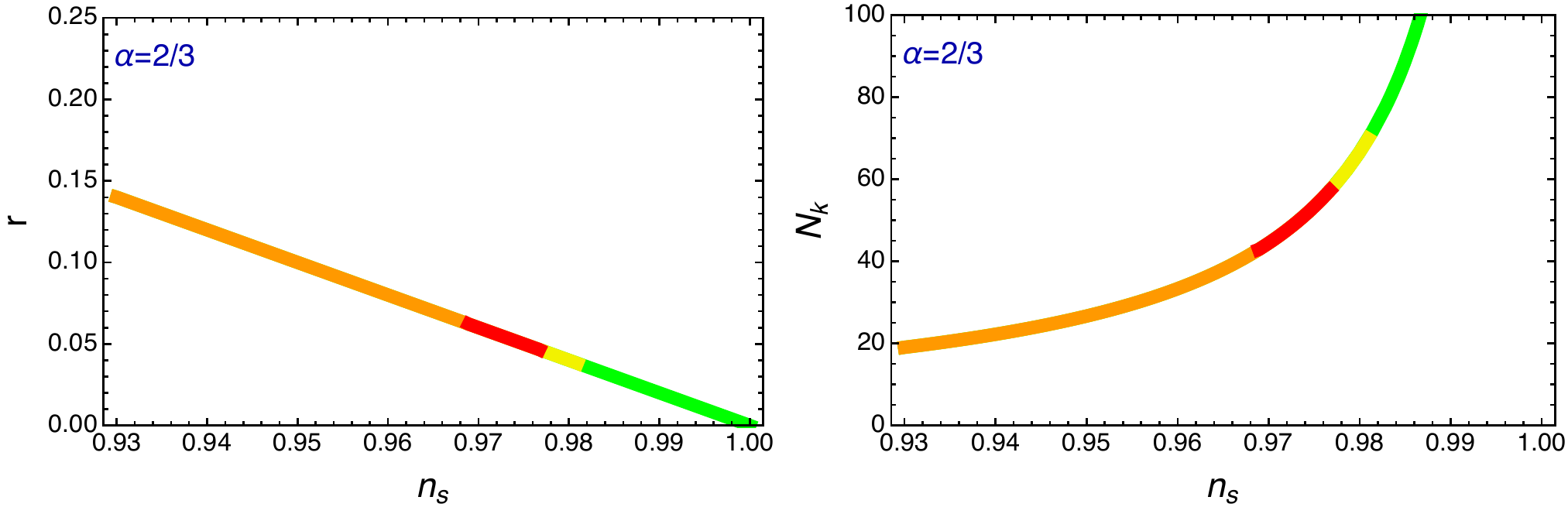}
\caption{Parameter space for $\phi^{2/3}$ inflation. The figures show $r$ and $N_k$ predictions that give the correct $A_s$ for the plotted $n_s$ at the pivot scale. The green portion of the line comprises the region of parameter space corresponding to reheating models with $w_{re}>1$, the yellow part corresponds to $w_{re} >1/3$, red to $w_{re} <1/3$ and orange to $w_{re}<0$. Note the most likely $w_{re}$, between $0$ and $1/3$, falls in the red region.}
    \label{fig:b}
\end{figure}

\begin{figure}[H]
\centering
    \includegraphics[width=14cm]{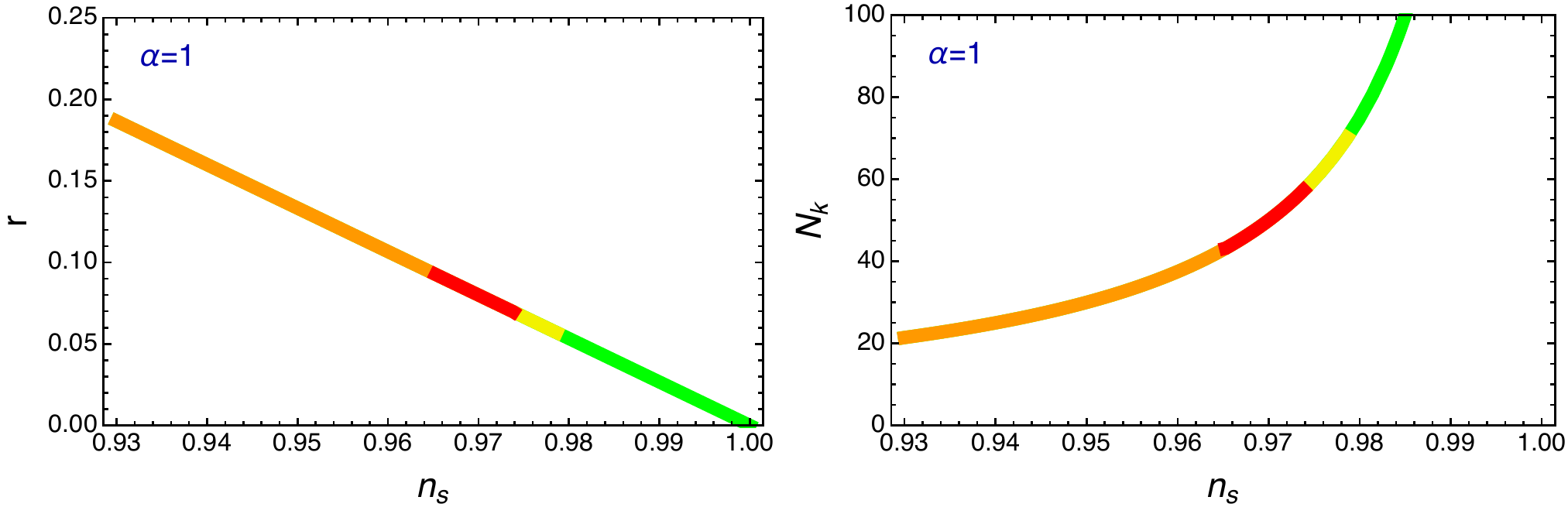}
\caption{Parameter space for $\phi$ inflation. Shading is as for Fig.~(\ref{fig:b}). }
    \label{fig:b1}
\end{figure}

\begin{figure}[H]
\centering
    \includegraphics[width=14cm]{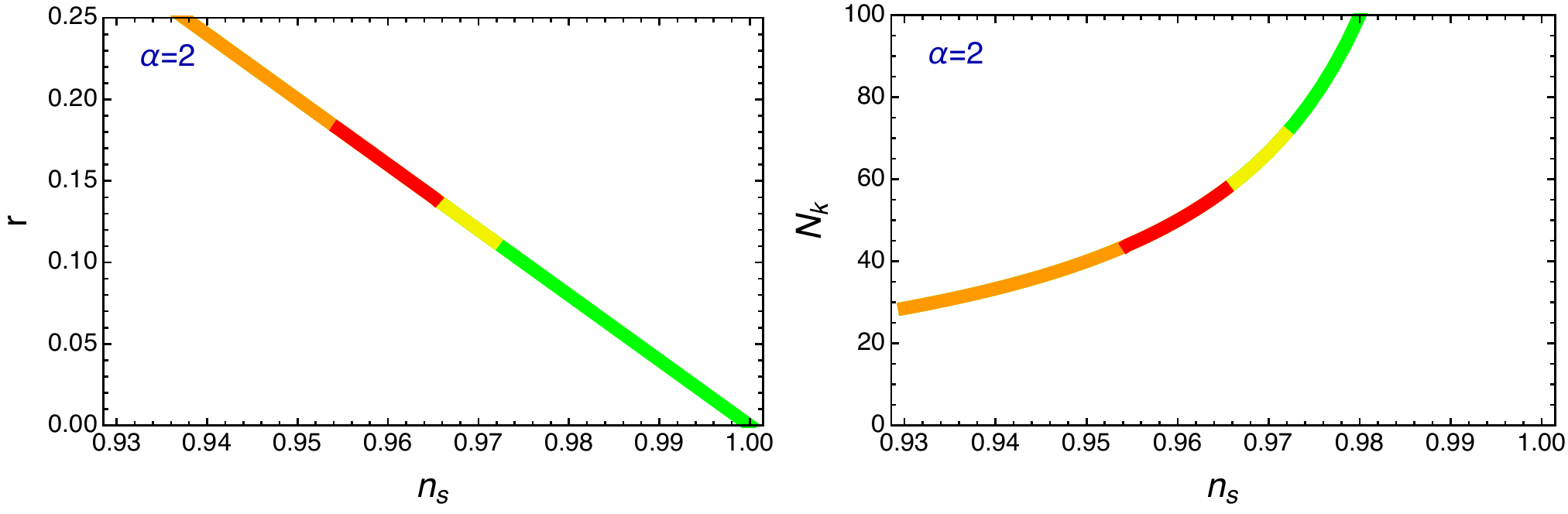}
\caption{Parameter space for $\phi^2$ inflation. Shading is as for Fig.~(\ref{fig:b}). }
    \label{fig:b2}
\end{figure}

Figs.~\ref{fig:b}-\ref{fig:b2} shows the parameter space in the $r$ and $N_k$ vs. $n_s$ plane, corresponding to the different reheating scenarios. We allow for any $N_k > 19$. We note again that there is no maximum allowed $N_k$. A minimum on $N_k$ is determined by the temperature at the end of reheating in order to solve the horizon and flatness problems. One finds that $N  > 24.9$ if reheating after inflation is to be above the BBN scale and $N >34.8$ for reheating above the electroweak scale in order for scales on the order of the horizon today ( i.e $l=2$) to have left the horizon during inflation. A simple estimate of the ratio of expansion scales between $l=2$, and Planck's pivot scale, at $l \approx 685.8$, if the expansion rate during Inflation were constant, is $\Delta N \approx \ln (l_2/l_1) \approx 6.5$. However, in the large field modes we are considering, the variation in $H$ is not negligible and the exact $\Delta N = \ln ( \frac{k_2\, H_1}{k_1\, H_2})$ is closer to $\Delta N \approx 5.9$. This means that for reheating greater than the BBN scale, one finds $N_k \geq 19$  (or $N_k \geq  29$  for reheating above the electroweak symmetry breaking scale). \\  

The green part of the line in Fig.~\ref{fig:b} corresponds to the region of parameter space that requires reheating models with $w_{re}$ larger than one, the yellow part corresponds to $w_{re} >1/3$, red to $w_{re} <1/3$ and orange to $w_{re}<0$. We stress that a value of $w_{re}$ between $0$ and $1/3$ is most likely and these solutions fall in the red band in Fig.~\ref{fig:b}.\\
One can see that requiring $0\leq w_{re} \leq 1/3$ corresponds to respectively setting an upper and a lower bound on the tensor-to-scalar ratio
\begin{equation}
\begin{cases}
0.05 \leq r \leq 0.06 \quad\quad  \text{for} \quad\alpha = \frac{2}{3}\,,\\
0.07 \leq r \leq 0.09  \quad\quad \text{for} \quad\alpha = 1\,,\\
0.14 \leq r \leq  0.18 \quad\quad \text{for} \quad\alpha = 2\,.\\
\end{cases}
\end{equation}
Since it now appears that the majority of BICEP2's signal is comprised of dust \cite{Ade:2015tva}, it is difficult to find a viable reheating scenario for $\phi^2$ inflation; if we loosen our restriction to just requiring $w_{re}<1$ then one obtains a bound $r\geq 0.11$, which is just inside the 2$\sigma$ limit \cite{Ade:2015tva}.  \\

The assumption $0\leq w_{re}\leq 1/3$ results in tighter constraints on $r$ than Planck's 2$\sigma$ bound on $n_{s}$ alone. For $\phi^2$, the $n_{s}$ 2$\sigma$ bound yields $0.08 \leq r \leq 0.18$. Restricting $w_{re}$ also provides stronger constraints on $N_{k}$: for $\phi^2$, the $n_{s}$ 2$\sigma$ bound yields $45 \leq N_{k}\leq 103$, whereas $0\leq w_{re}\leq 1/3$ yields $44 \leq N_{k}\leq 57$.


\section{Starobinsky model}
\label{sec4}

The action for the Starobinsky model \cite{Starobinsky:1980te} has the form
\begin{align}\label{seq1}
S = \int d^4 x \sqrt{-g} \left[\frac{M_P^2}{2} (R + \alpha R^2) + \mathcal{L}_{matter} \right],
\end{align}
where $R$ is the Ricci scalar. Performing a conformal transformation \cite{Wetterich:1987fk,Kalara:1990ar}
\begin{align}
\tilde{g}_{\mu \nu} = \omega^2 g_{\mu \nu},
\end{align}
where $\omega^2 = 1 + 2 \alpha R$, the action (\ref{seq1}) is rewritten as the canonical Einstein-Hilbert action plus other terms which form a modified $\mathcal{L}_{matter}$
\begin{align}
S = \int d^4 x \sqrt{- \tilde{g}} \left[\frac{M_P^2}{2} \left[ \tilde{R} - \frac{\alpha \phi^2}{(1+ 2 \alpha \phi)^2} - \frac{6 \alpha^2}{(1+ 2 \alpha \phi)^2} (\tilde{\partial} \phi)^2 \right] + \mathcal{L}_{matter} \right],
\end{align}
where what we now call $\phi$ is equal to $R$, the original, untransformed Ricci scalar. Notice that $\tilde{\partial}^{\alpha}$ carries factors of the metric, therefore $\neq \partial^{\alpha}$.  Next one defines $\bar{\phi}$, a canonically normalized version of $\phi$ 
\begin{align}
\bar{\phi} = \sqrt{\frac{3}{2}} M_P \ln(1+ 2 \alpha \phi).
\end{align}
Rewriting the action in terms of $\bar{\phi}$ one finds
\begin{align}
S = \int d^4 x \sqrt{- \tilde{g}} \left[\frac{M_P^2}{2} \left[\tilde{R} - \frac{1}{4 \alpha} \left(1 - e^{- \sqrt{\frac{2}{3}} \frac{\bar{\phi}}{M_P}}  \right)^2  \right] - \frac{1}{2} (\tilde{\partial} \bar{\phi})^2 + e^{- 2 \sqrt{\frac{2}{3}} \frac{\bar{\phi}}{M_P}} \mathcal{L}_{matter}   \right].
\end{align}
If one assumes that the other fields in $\mathcal{L}_{matter}$ are subdominant during inflation and can be ignored, then one can verify that this Einstein frame action behaves as normal gravity plus a canonical scalar field with the potential
\begin{align}
V = \frac{M_P^2}{8 \alpha}  \left(1- e^{- \sqrt{\frac{2}{3}} \frac{\bar{\phi}}{M_P}} \right)^2\,.
\end{align}
Dropping the bar on $\phi$ from now on, but continuing to work with the canonical version of the field, one can easily compute the number of e-foldings between the horizon exit of the pivot scale and the end of inflation 
\begin{align}
N_k = \frac{1}{M_P^2} \int_{\phi_{end}}^{\phi_k} \frac{V}{V'}\, d \phi=\frac{1}{2 M_P^2} \sqrt{\frac{3}{2}} \left[M_P \sqrt{\frac{3}{2}} e^{\sqrt{\frac{2}{3}} \frac{\phi}{M_P}} - \phi  \right] \Big|^{\phi_k}_{\phi_{end}}.
\end{align}
With the approximations $\phi_k \gg \phi_{end}$, and $M_P  e^{\sqrt{\frac{2}{3}} \frac{\phi_k}{M_P}} \gg \phi_k$ , the previous expression simplifies to
\begin{align}
N_k = \frac{3}{4} e^{\sqrt{\frac{2}{3}} \frac{\phi_k}{M_P} }
\end{align}
which can be inverted for $\phi_k$
\begin{align}\label{eq9}
\phi_k = \sqrt{\frac{3}{2}} M_P \ln \left(\frac{4}{3} N_k \right)
\end{align}
The next step is to compute $\epsilon_k$ and $\eta_k$ in order to derive $N_k$ as a function of $n_s$ using $n_s = 1 - 6 \epsilon + 2 \eta$. The slow-roll parameters have the following form
\begin{align}\label{eq33}
\epsilon_k \simeq  \frac{3}{4 N_k^2},\quad\quad\quad \eta_k = - \frac{1}{N_k},
\end{align}
where Eq.~(\ref{eq9}) was used along with the approximation $N_{k}\gg 1$. From Eq.~(\ref{eq33}) then one finds 
\begin{align}\label{n1}
N_k = \frac{2}{1 - n_s}.
\end{align}
Using the expressions above, one derives $H_k$ as a function of $n_s$ and $A_s$
\begin{eqnarray}\label{n2}
& &H_k = \pi M_P  \sqrt{\frac{3}{2} A_s} (1- n_s),\\\label{n3}
& &V_{end} = \frac{9}{2} \pi^2 M_P^4 A_s (1-ns)^2 \frac{ \left(\frac{1}{\frac{\sqrt{3}}{2}+1} \right)^2}{ \left(1 - \frac{3}{8}(1-n_s) \right)^2}
\end{eqnarray}
Eqs.~(\ref{n1})-(\ref{n3}) are all that is needed to derive the results for the duration and for the temperature of reheating.

\begin{figure}[H]
\centering
    \includegraphics[width=9cm]{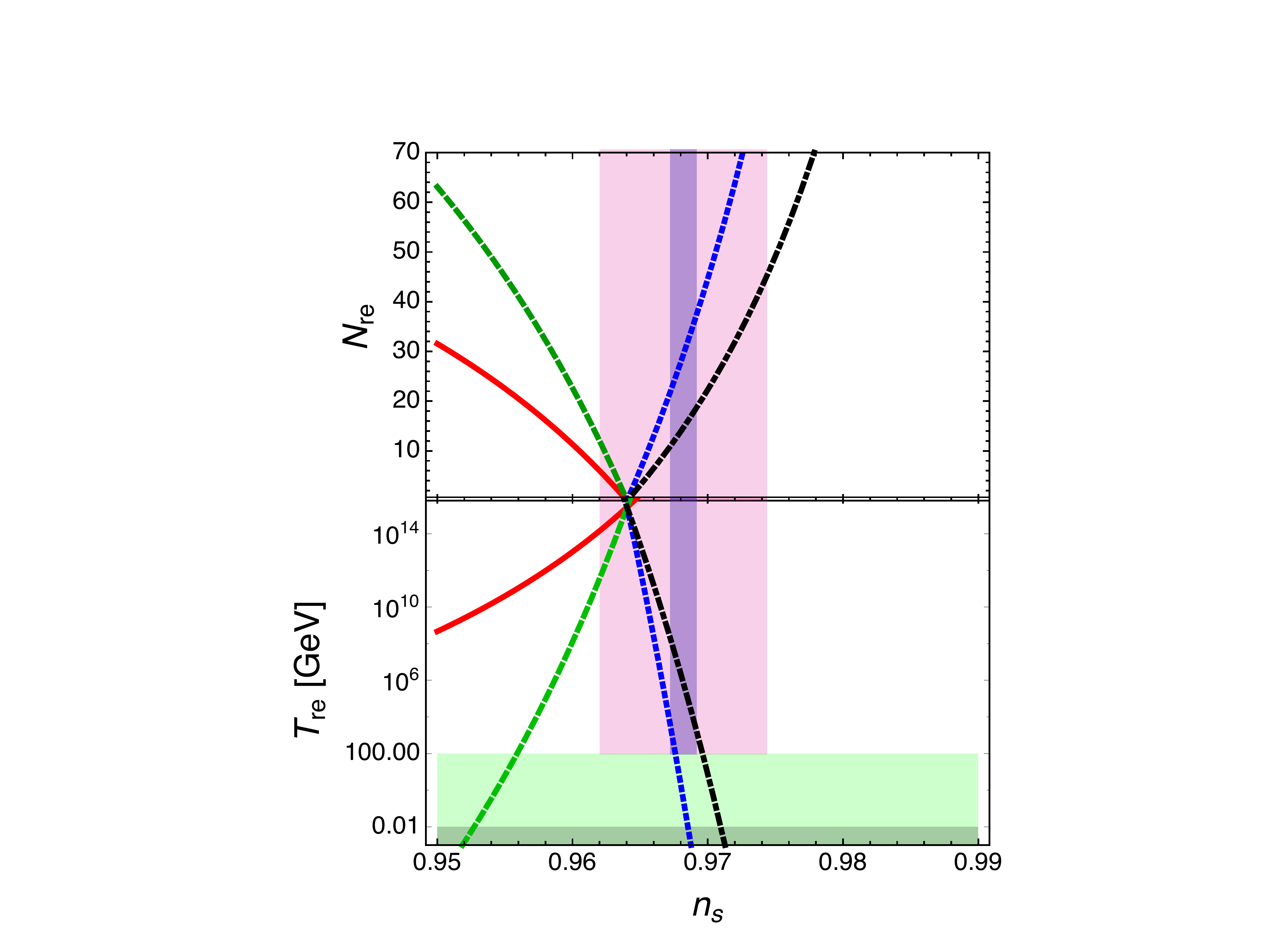}
\caption{Plots of $N_{re}$ and $T_{re}$, the length of reheating and the temperature at the end of reheating respectively, for Starobinsky and Higgs inflation.  All curves and shaded regions are as for Fig.~\ref{fig:a}}
    \label{fig:c}
\end{figure}

Fig.~\ref{fig:c} shows good compatibility with Planck's $1 \sigma$ bounds on $n_s$ for all the possible $w_{re}$ values. Also, if one does not put any restrictions on the value of $w_{re}$ then any temperature between the BBN bound and the instantaneous reheating value is allowed within the 1$\sigma$ bound.

\begin{figure}[H]
\centering
   \includegraphics[width=14cm]{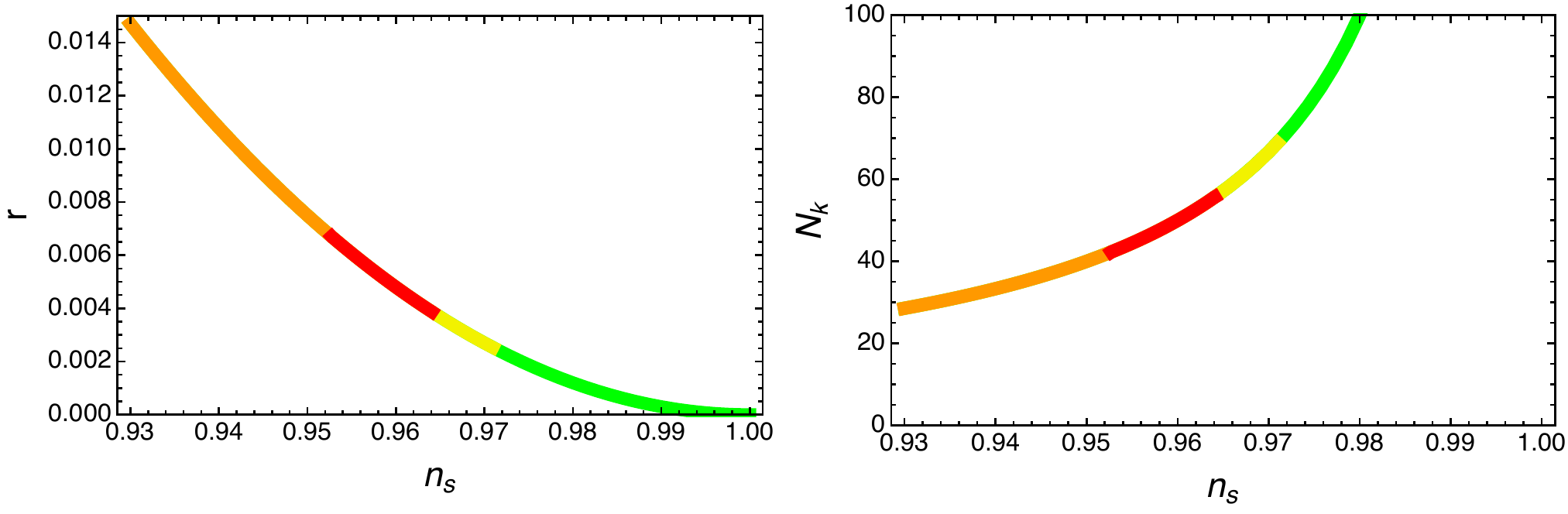}
\caption{Parameter space for Starobinsky inflation. Shading is as for Fig.~\ref{fig:b}. }
    \label{figd}
\end{figure}

\noindent We plot in Fig.~\ref{figd} the parameter space in the $r$ and $N_k$ vs. $n_s$ plane for Starobinsky inflation, for different ranges of $w_{re}$\footnote{{See also \cite{Martin:2014vha,Martin:2014nya} for $n_{s}$ vs $r$ plots in the Starobinsky model for $w_{re}=0$.}}. For $0 <w_{re} <1/3$, the corresponding range for the spectral index is $0.953 < n_s < 0.964$. This also corresponds to the range $0.004 \leq r \leq 0.007$ and $42 \leq N_k \leq 56$.

\section{Higgs Inflation}
\label{sec5}
The idea behind Higgs inflation \cite{Bezrukov:2007ep} is to allow the Standard Model Higgs field to be the inflaton by adding a non-minimal coupling to gravity. The Jordan frame action is
\begin{align}
S = \int d^4 x \sqrt{-g} \left[\frac{M_P^2}{2} R \left(1 + 2 \xi \frac{H^{\dagger} H}{M_P^2} \right) + \mathcal{L}_{matter} \right]\,,
\end{align}
where H is the Higgs doublet. We may again perform a conformal transformation to write the action in the form of Einstein gravity plus a modified $\mathcal{L}_{matter}$. The transformation is given by $\tilde{g}_{\mu \nu} = \omega^2 g_{\mu \nu}$ with $\omega^2 = 1 + 2 \xi \frac{H^{\dagger} H}{M_P^2}$. Rewriting the action in terms of the transformed metric, we find
\begin{align}
S = \int d^4 x \sqrt{- \tilde{g}}  \left[\frac{M_P^2}{2} \tilde{R} - \frac{3 \xi^2}{\omega^4 M_P^2}  \left(\tilde{\partial} H^{\dagger} H \right)^2 + \frac{1}{\omega^4} \mathcal{L}_{matter} \right]\,.
\end{align}
Next, one extracts the kinetic and potential terms for the Higgs field contained within $\mathcal{L}_{matter}$. One can use $V_h = \frac{\lambda}{4} (H^{\dagger} H - \frac{\nu^2}{2})^2$, dropping the $\nu$ part (we are interested in inflation scales much larger than electroweak scale). Ignoring all the Higgs interactions with other fields, and only considering its self coupling (which we assume is the dominant term in the Higgs potential at inflation scales)
\begin{align}
S = \int d^4 x \sqrt{- \tilde{g}} \left[\frac{M_P^2}{2} \tilde{R} - \frac{3 \xi^2}{\omega^4 M_P^2}  \left(\tilde{\partial} H^{\dagger} H \right)^2  - \frac{1}{\omega^4} \left(\partial H^{\dagger} \right)^2- \frac{\lambda}{4 \omega^4} \left (H^{\dagger} H \right)^2 + \frac{1}{\omega^4} \mathcal{L}_{matter} \right]\,,
\end{align}
where now $\mathcal{L}_{matter}$ comprises all the matter fields except the Higgs. Note: one needs to convert $\partial^{\alpha} \rightarrow \omega^2 \tilde{\partial}^{\alpha}$.  The Higgs is no longer canonical because of the effect of the non-minimal coupling. To canonically normalize all four of the Higgs degrees of freedom, one must work in unitary gauge, where three of the four degrees of freedom are equal to zero, $H = \frac{1}{\sqrt{2}} \left( \begin{array}{cc}
0  \\
h \end{array} \right)$
\begin{align}
S = \int d^4 x \sqrt{- \tilde{g}} \left[\frac{M_P^2}{2} \tilde{R} - \frac{3 \xi^2 h^2}{\omega^4 M_P^2} \left(\tilde{\partial} h \right)^2 - \frac{1}{2 \omega^2} \left(\tilde{\partial} h \right)^2 - \frac{\lambda}{4 \omega^4} h^4 + \frac{1}{\omega^4} \mathcal{L}_{matter} \right]\,.
\end{align}
The canonically normalized version of $h$ is $\bar{h}$, defined as
\begin{align}\label{app1}
\frac{\partial \bar{h}}{\partial h} = \frac{1}{ \left(1 + \frac{\xi}{M_P^2} h^2 \right)} \sqrt{1 + \frac{\xi}{M_P^2}  h^2 (6 \xi + 1)  }.
\end{align}
Before integrating the previous equation, it is useful to introduce a few approximations. First one uses $6 \xi \gg 1$. To get a successful inflation model, one should require $\xi \approx 10^4$. Next one uses the condition $(6 \xi)/(M_P^2)  h^2 \gg 1$. $h \approx M_P$ when inflation ends, and therefore $h> M_P$ for the duration of inflation. This allows one to rewrite Eq.~(\ref{app1}) as 
\begin{align}
\bar{h}= \frac{\sqrt{6} \xi}{M_P} \int dh\, \frac{h}{1 + \frac{\xi h^2}{M_P^2}}\,,
\end{align}
which integrates to
\begin{align}
\bar{h}= \sqrt{\frac{3}{2}} M_P \ln \left(1+ \frac{\xi h^2}{M_P^2} \right).
\end{align}
Rewriting the action in terms of $\bar{h}$, one finds
\begin{align}
S = \int d^4 x \sqrt{- \tilde{g}} \left[\frac{M_P^2}{2} \tilde{R} - \frac{1}{2 }  \left(\tilde{\partial} \bar{h} \right)^2 - \frac{\lambda M_P^4}{4 \xi^2}  \left(1 - e^{- \sqrt{\frac{2}{3}} \frac{\bar{h}}{M_P}  } \right)^2 + e^{-2 \sqrt{\frac{2}{3}} \frac{\bar{h}}{M_P} } \mathcal{L}_{matter} \right]\,, 
\end{align}
The potential term for the canonical field takes the same form as the Starobinsky potential with the identification $(1)/(8 \alpha) = (\lambda M_P^2)(4 \xi^2)$. Since we have a canonical field evolving in the same potential as the Starobinsky case, the Higgs inflation model gives the same predictions for $N_{re}$ and $T_{re}$ (see also \cite{Martin:2014vha,Martin:2014nya}).  We note Starobinsky and Higgs inflation have different low scale behavior 
\cite{Bezrukov:2008ut,GarciaBellido:2008ab,Bezrukov:2011gp,Gorbunov:2012ns,Kehagias:2013mya}
 and so while the allowed parameter space as a function of $w_{re}$ is the same, the $w_{re}$ that is most likely for Starobinsky vs. Higgs inflation is likely to differ. Tighter constraints could be obtained by considering gravitational, Planck suppressed couplings in the Starobinsky case, and standard model couplings in the Higgs case \cite{Bezrukov:2008ut,GarciaBellido:2008ab,Bezrukov:2011gp}. Of course new physics may modify the running of the couplings or add new couplings at these high scales (see for example \cite{Gorbunov:2012ns}); in this respect, our approach of characterizing an allowed parameter space by assuming a range of $w_{re}$ between 0 and 1/3 can usefully help bracket different allowed scenarios.}


\section{Natural Inflation}
\label{sec6}

The potential for natural inflation is \cite{Freese:1990rb}
\ba
V(\phi)=\Lambda^{4}\left[1+\cos\left(\frac{\phi}{f}\right)\right]\,.
\ea
The number of e-folds $N_{k}$ between the time the pivot scale modes crossed outside the horizon and the end of inflation is given by
\ba\label{ef}
N_{k}=\left(\frac{f}{M_{P}}\right)^{2}\ln\left[\frac{\sin^{2}\left(\chi_{end}/2 \right)}{\sin^{2}\left(\chi_{in}/2\right)}\right]
\ea
where $\chi\equiv\phi/f$. The slow-roll parameters have the following form
\ba\label{ep}
\epsilon=\frac{1}{2}\left(\frac{M_{P}}{f}\right)^{2}\left[\frac{1-\cos(\chi)}{1+\cos(\chi)}\right]\,,\quad\quad\quad \eta=-\left(\frac{M_{P}}{f}\right)^{2}\frac{\cos\chi}{1+\cos\chi}.
\ea
The field value at the end of inflation can be determined by setting $\epsilon=1$; this leads to the following equation for $\chi_{end}$
\ba
\frac{1}{2}\frac{M_{P}^{2}}{f^{2}}\frac{\sin^{2}\left(\chi_{end}\right)}{\left[1+\cos\left(\chi_{end}\right)\right]^{2}}=1.
\ea
The solution is
\ba
\cos(\chi_{end})=\frac{-1+b}{1+b}\,,\quad\quad\quad\quad b\equiv\left(\frac{M_{P}}{\sqrt{2}f}\right)^{2},
\ea
The number of e-folds in Eq.~(\ref{ef}) can be written as 
\ba
N_{k}=\left(\frac{f}{M_{P}}\right)^{2}\ln\left[\frac{1-\cos(\chi_{end})}{1-\cos(\chi_{in})}\right]=\left(\frac{f}{M_{P}}\right)^{2}\ln\left[\frac{2}{(1+b)}\frac{1}{(1-\cos(\chi_{in}))}\right].
\ea
The value of the field at the pivot scale during inflation is then given by
\ba\label{chi}
\cos(\chi_{in})=1-z\,,\quad\quad\quad\quad z\equiv \frac{2}{(1+b)}\exp{\left[-N_{k}\left(\frac{M_{P}}{f}\right)^{2}\right]}.
\ea
Using (\ref{ep}) and (\ref{chi}), one finds
\ba
n_{s}-1\equiv -6\epsilon+2\eta=-\left(\frac{M_{P}}{f}\right)^{2}\left(\frac{2+z}{2-z}\right),
\ea
which leads to
\ba
N_{k}=-\left(\frac{f}{M_{P}}\right)^{2}\ln\left[\left(1+\frac{M_{P}^{2}}{2f^{2}}\right)\left(\frac{(1-n_{s})-\frac{M_{P}^{2}}{f^{2}}}{(1-n_{s})+\frac{M_{P}^{2}}{f^{2}}}\right)\right].
\ea
Notice that the previous expression is positive and real only if the argument of the logarithm is defined between zero and one
\ba\label{cond2}
0 < \left(1+\frac{M_{P}^{2}}{2f^{2}}\right)\left(\frac{(1-n_{s})-\frac{M_{P}^{2}}{f^{2}}}{(1-n_{s})+\frac{M_{P}^{2}}{f^{2}}}\right)< 1.
\ea
The conditions (\ref{cond2}) are equivalent to requiring that
\ba\label{cond1}
\left(\frac{f}{M_{P}}\right)^{2}>\frac{1}{(1-n_{s})}  \quad\quad \text{and}   \quad\quad 3+n_{s}+\left(\frac{M_{P}}{f}\right)^{2} > 0 .
\ea
The second condition in (\ref{cond1}) is always true.  The first condition implies a minimum $f$ for each $n_s$, and the bound on $f$ increases with increasing $n_s$. Using the central value for the Planck constraints on the spectral index, then (\ref{cond1}) gives $f>5.6 \,M_{P}$. \\ 

\noindent The tensor-to-scalar ratio can be expressed in terms of $n_{s}$
\ba
r=16\,\epsilon = 4\left[(1-n_{s})-\frac{M_{P}^{2}}{f^{2}}\right].
\ea
Using $V_{in}=\Lambda^{4}\left[1+\cos(\chi_{in})\right]\simeq 3H^{2}M_{P}^{2}$,
one finds
\ba
V_{end}=3H^{2}M_{P}^{2}\left[\frac{1+(1-n_{s})\left(\frac{f}{M_{P}}\right)^{2}}{2+4\left(\frac{f}{M_{P}}\right)^{2}}\right].
\ea

\begin{figure}[H]
\centering
    \includegraphics[width=16cm]{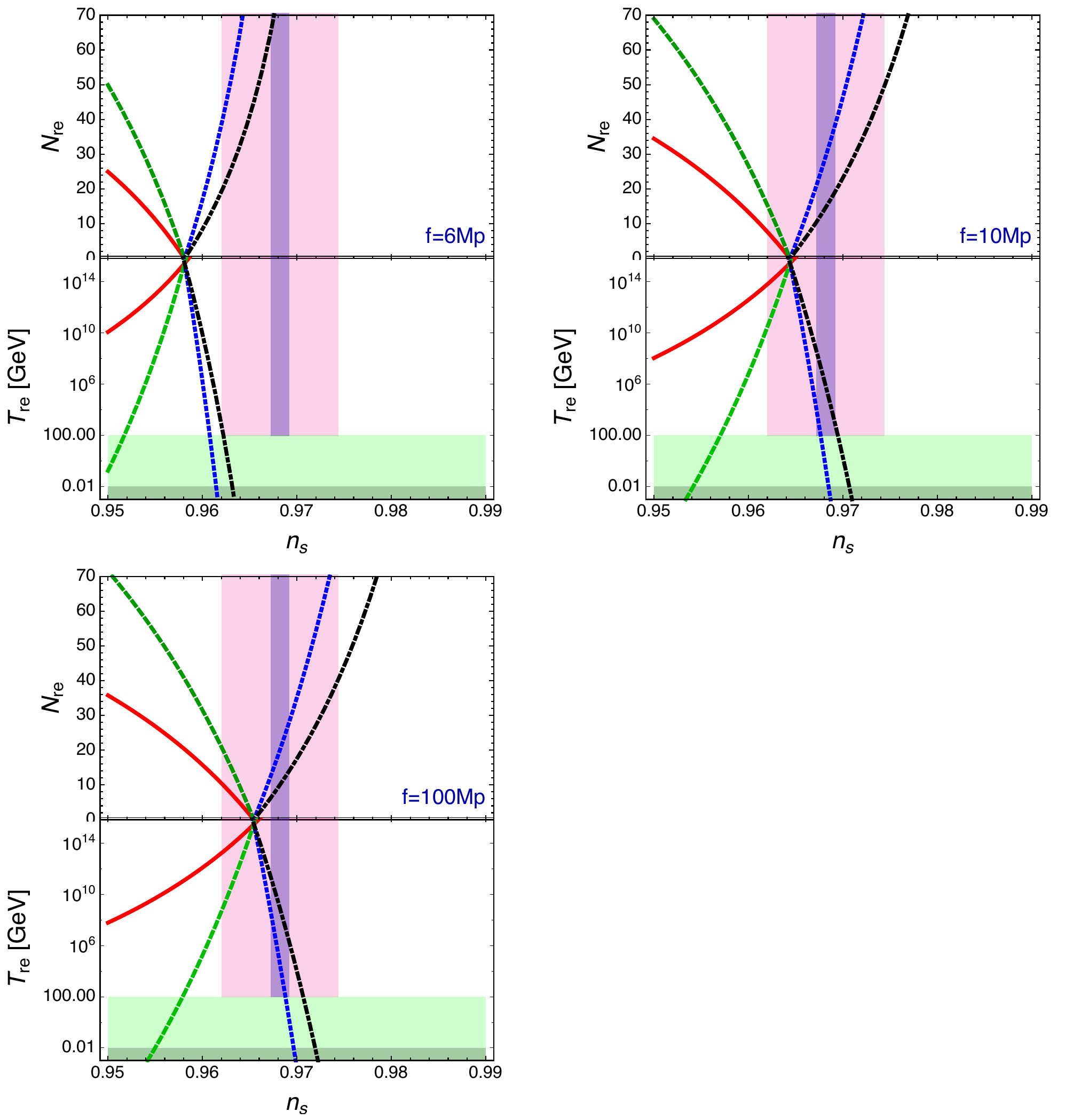}
\caption{We show $N_{re}$ and $T_{re}$, the length of reheating and the temperature at the end of reheating respectively, for natural inflation, for 3 values of the coupling $f$. Again, shading is as in Fig.~\ref{fig:a} }
    \label{fig:e}
\end{figure}

Fig.~\ref{fig:e} shows $N_{re}$ and $T_{re}$ solutions for various reheating parameters $w_{re}$ and for various couplings $f$ in natural inflation.  Unlike polynomial inflation, or Starobinsky/ Higgs inflation, natural inflation has an extra free parameter, and so one no longer gets a precise prediction for the temperature and length of reheating once a reheating model, $w_{re}$, and $n_s$ are specified. But one can get reasonable bounds on the coupling $f$ such that a viable reheating model exists.\\

\begin{figure}[H]
\centering
    \includegraphics[width=14cm]{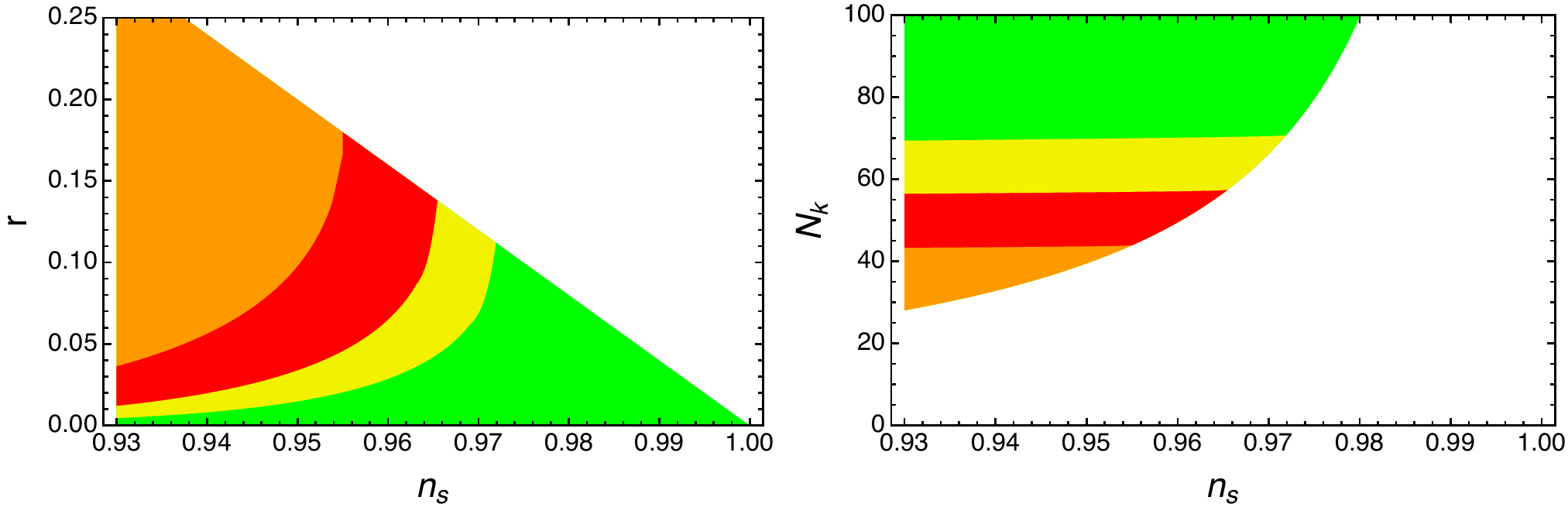}
\caption{Representation of the $r$ and $N_k$ vs. $n_s$ plane for natural inflation. The colored region is the entire allowed parameter space that can produce the measured $A_s$ value at the pivot scale, and the $n_s$ value at the pivot scale as plotted. Shading again follows Fig.~\ref{fig:b}}
    \label{fig:f}
\end{figure}

One can obtain separate, and stronger constraints on $f$ based on the requirement for viable reheating.  These constraints likewise are functions of $n_s$, and their effects are displayed in Figure \ref{fig:f}\footnote{{See also \cite{Martin:2014vha,Martin:2014nya} for $n_s$ vs $r$ plots in the natural inflation model for $w_{re}=0$.}}. \\

There is no upper limit on $f$. For $f \gtrsim 14 M_P$ the various $w_{re}$ lines reach an asymptotic form.  As a result even for very large $f$, there is a valid solution for each $w_{re}$ value consistent with Planck's $1 \sigma$ bounds. The asymptotic solution for large $f$ for $0\leq w_{re} \leq 1/3$ corresponds to a solution for the spectral index in the range $0.956\leq n_{s}\leq 0.965$. \\ 

Note the minimum on $f$ increases with increasing $n_s$, such that for larger $n_s$, a larger $f$ is needed to find a solution consistent with the reheating model being considered. The top limit in Fig.~\ref{fig:f} A (the bottom limit in Fig.~\ref{fig:f} B) is approached asymptotically for large $f$.  The bottom part of the parameter space in Fig.~\ref{fig:f} A (top part in Fig.~\ref{fig:f} B) corresponds to small $f$. Everywhere in the figure $N_k > 19$, such that inflation lasts long enough to allow for BBN.\\

Using Planck's $2 \sigma$ bounds on $n_s$, requiring $w_{re} \leq 1$ gives $r \geq 0.02$, and requiring $w_{re} \leq 1/3$ gives $r \geq 0.05$.  For $w_{re}\leq 1/3$, values of $n_{s}$ smaller than Planck's central value would be favored for any $f$. The weakest constraint is for large values $f$, for which $n_{s}\leq 0.965$.


\section{Hilltop inflation}
\label{sec7}

The potential is given by \cite{Linde:1981mu,Linde:1984cd}
\ba
V\left(\phi\right)=M^{4}\left[1-\left(\frac{\phi}{\mu}\right)^{p}\right].
\ea

We begin by considering $p > 2$. The exact expression for the number of e-foldings between the time the pivot scale crossed outside the horizon and the end of inflation is
\ba\label{N}
N_{k}=\frac{\mu^{2}}{2p M_{P}^{2}}\left[\chi_{in}^{2}-\chi_{end}^{2}+\frac{2}{p-2}\chi_{in}^{2-p}-\frac{2}{p-2}\chi^{2-p}_{end}\right].
\ea
where one defines $\chi\equiv\phi/\mu$.\\
The slow-roll parameters are
\ba\label{epsilon}
\epsilon=\frac{p^{2}}{2}\frac{M_{P}^{2}}{\mu^{2}}\frac{\chi^{2(p-1)}}{\left(1-\chi^{p}\right)^{2}}\,,\quad\quad\eta\equiv -p(p-1)  \frac{M_{P}^{2}}{\mu^{2}} \frac{\chi^{p-2}}{\left(1-\chi^{p}\right)}. 
\ea
Setting $\epsilon=1$ at the end of inflation one derives the equation for $\chi_{end}$
\ba\label{chiend}
\frac{p^{2}}{2}\frac{M_{P}^{2}}{\mu^{2}}\frac{\chi_{end}^{2(p-1)}}{\left(1-\chi_{end}^{p}\right)^{2}}=1.
\ea
Let us consider the case where $\mu > M_{P}$ and define $q\equiv M_{P}/\mu$. For small values of $q$, one can search for a solution for $\chi_{end}$ in the form of a Taylor expansion around $q=0$
\ba
\chi_{end}=a_{0}+a_{1}\, q+\frac{1}{2}\,a_{2} \,q^{2}+\mathcal{O}(q^{3}).
\ea
One can show that, up to order $q^{2}$, a solution to Eq.~(\ref{chiend}) is (see e.g. \cite{Martin:2014vha})
\ba\label{solchiend}
\chi_{end}=1-\frac{1}{\sqrt{2}}\,q+\frac{(p-1)}{4}\,q^{2}.
\ea
Similarly, one can look for a solution for the initial value of the scalar field using (\ref{solchiend}) and (\ref{N}), to find\ba
\chi_{in}=1-\sqrt{\frac{1+4\,N_{k}}{2}}\,q +\mathcal{O}(q^{2}).
\ea
Using (\ref{solchiend}) in the expression for the slow-roll parameters, (\ref{epsilon}), the spectral index as a function of $N_{k}$  is
\ba\label{ns}
n_{s}-1\simeq -\frac{6}{1+4 N_{k}}, \quad\quad\quad\longrightarrow\quad\quad\quad N_{k}\simeq \frac{1}{4}\left(\frac{6}{1-n_{s}}-1\right).
\ea
The tensor-to-scalar ratio is
\ba\label{r}
r\simeq \frac{8}{3}(1-n_{s}).
\ea
Notice that (\ref{ns}) and(\ref{r}) only apply for small values of $q$, more precisely for
\ba
q < \frac{\sqrt{2}}{(p-1)\sqrt{1+4 \,N_{k}}}.
\ea
For $p\in (3,8)$ and for $N_{k}\in (30,100)$, the previous condition is satisfied if $q\leq 0.01$.\\
Within the same range of validity, the potential at the end of inflation is given by
\ba\label{pot}
V_{end}\simeq \sqrt{\frac{3}{2}}H^{2}M_{P}^{2}\sqrt{1-n_{s}}.
\ea
For $p\leq 2$ one derives the same results as Eqs.~(\ref{ns})-(\ref{r}) and (\ref{pot}).   For $p=2$, however, a new expression for $N_k$ is required. In this case we find
\ba\label{N2}
N_{k}=\frac{\mu^{2}}{2 M_{P}^{2}}\left[\frac{\chi_{in}^{2}}{2}-\frac{\chi_{end}^{2}}{2} - \ln \chi_{in} + \ln \chi_{end}  \right].
\ea

\begin{figure}[H]
\centering
   \includegraphics[width=16cm]{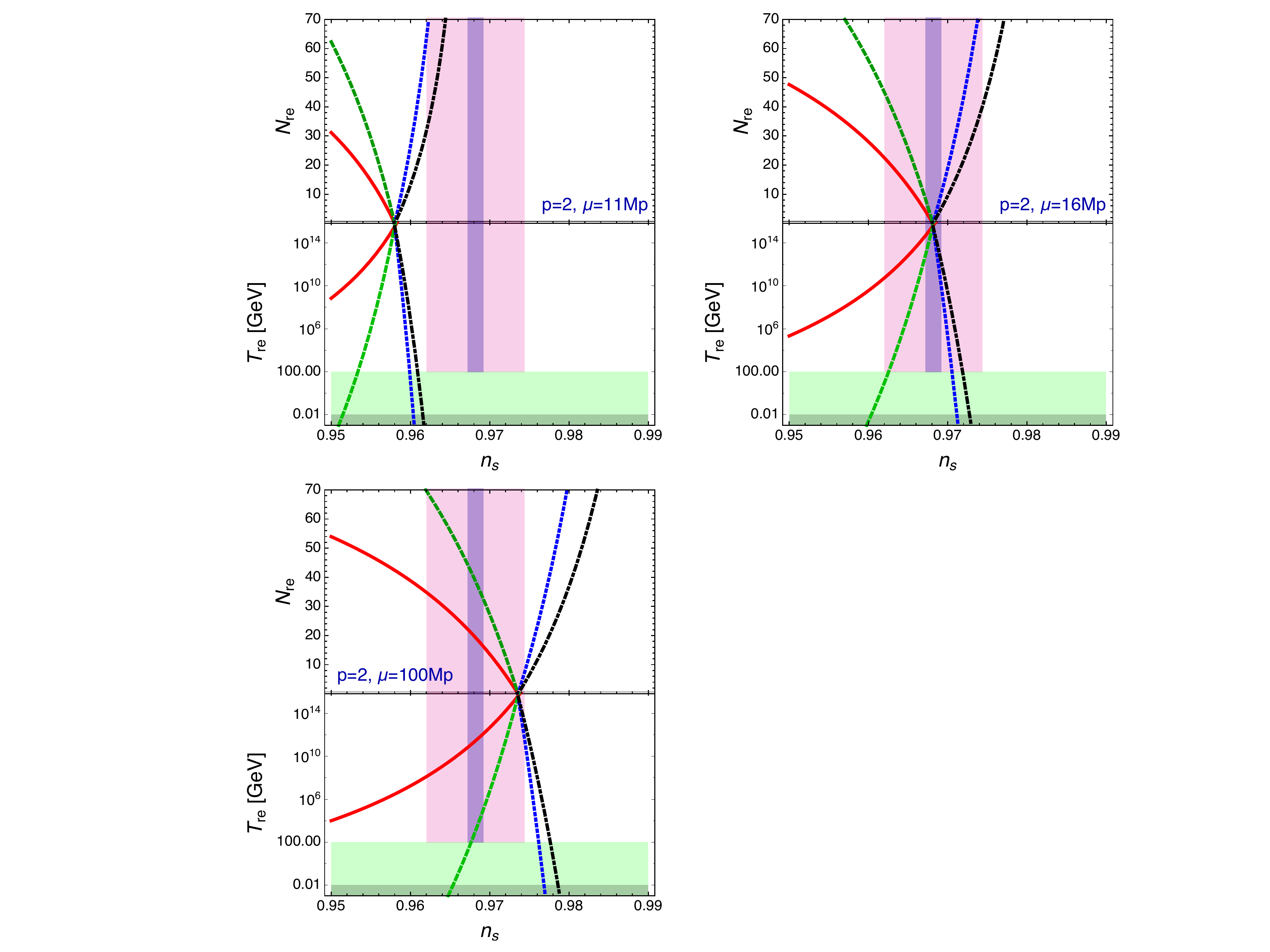}
\caption{Plots of $N_{re}$ and $T_{re}$, for hilltop inflation with $p=2$ and for three different values of $\mu$. Shading is as in Fig.~\ref{fig:a}
}
   \label{fig:g}
\end{figure}

\begin{figure}
\centering
    \includegraphics[width=16cm]{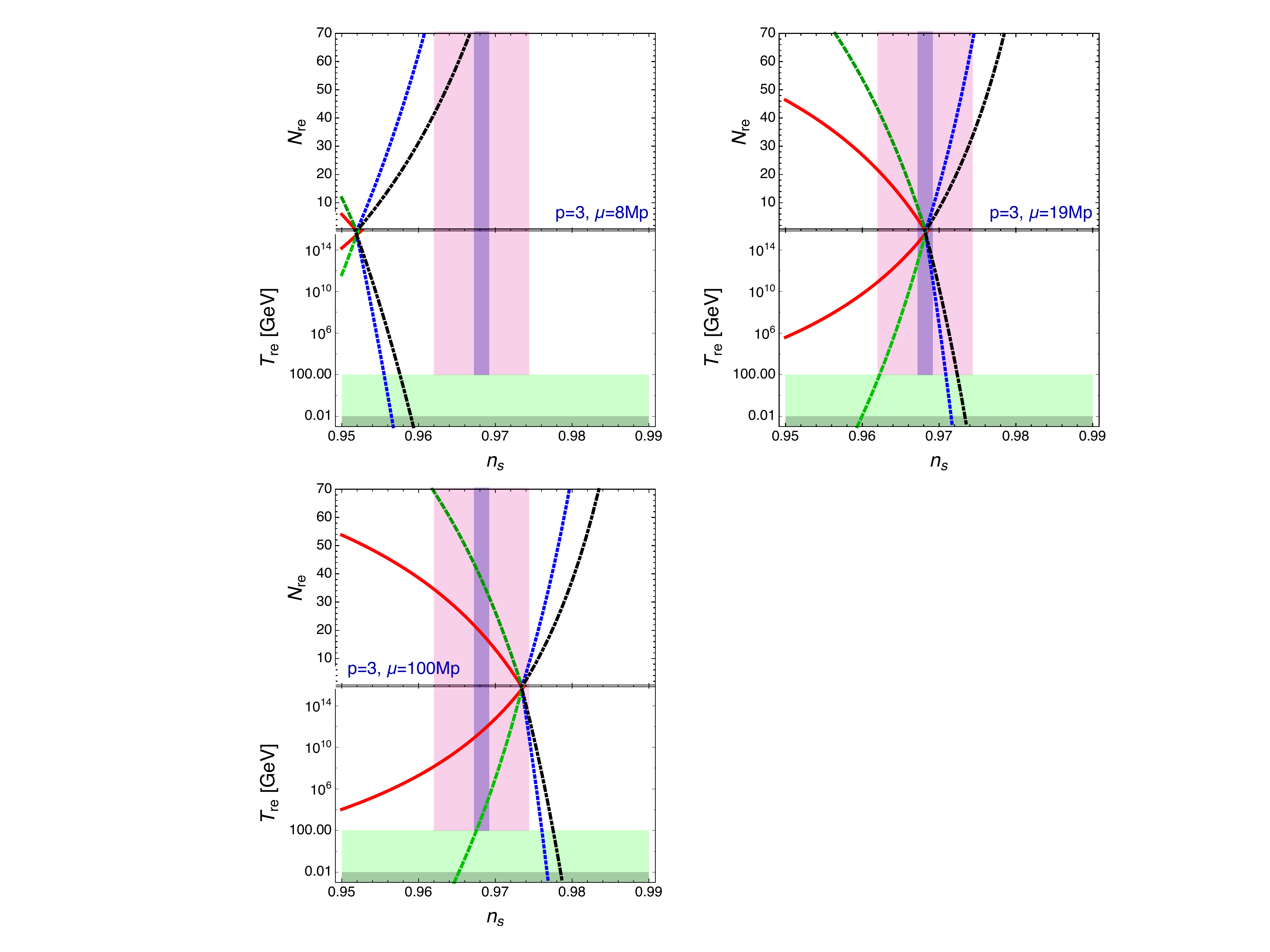}
\caption{Plots of $N_{re}$ and $T_{re}$, for hilltop inflation with $p=3$ and for three different values of $\mu$. Shading is as in Fig.~\ref{fig:a}
}
    \label{fig:h}
\end{figure}

\begin{figure}
\centering
    \includegraphics[width=16cm]{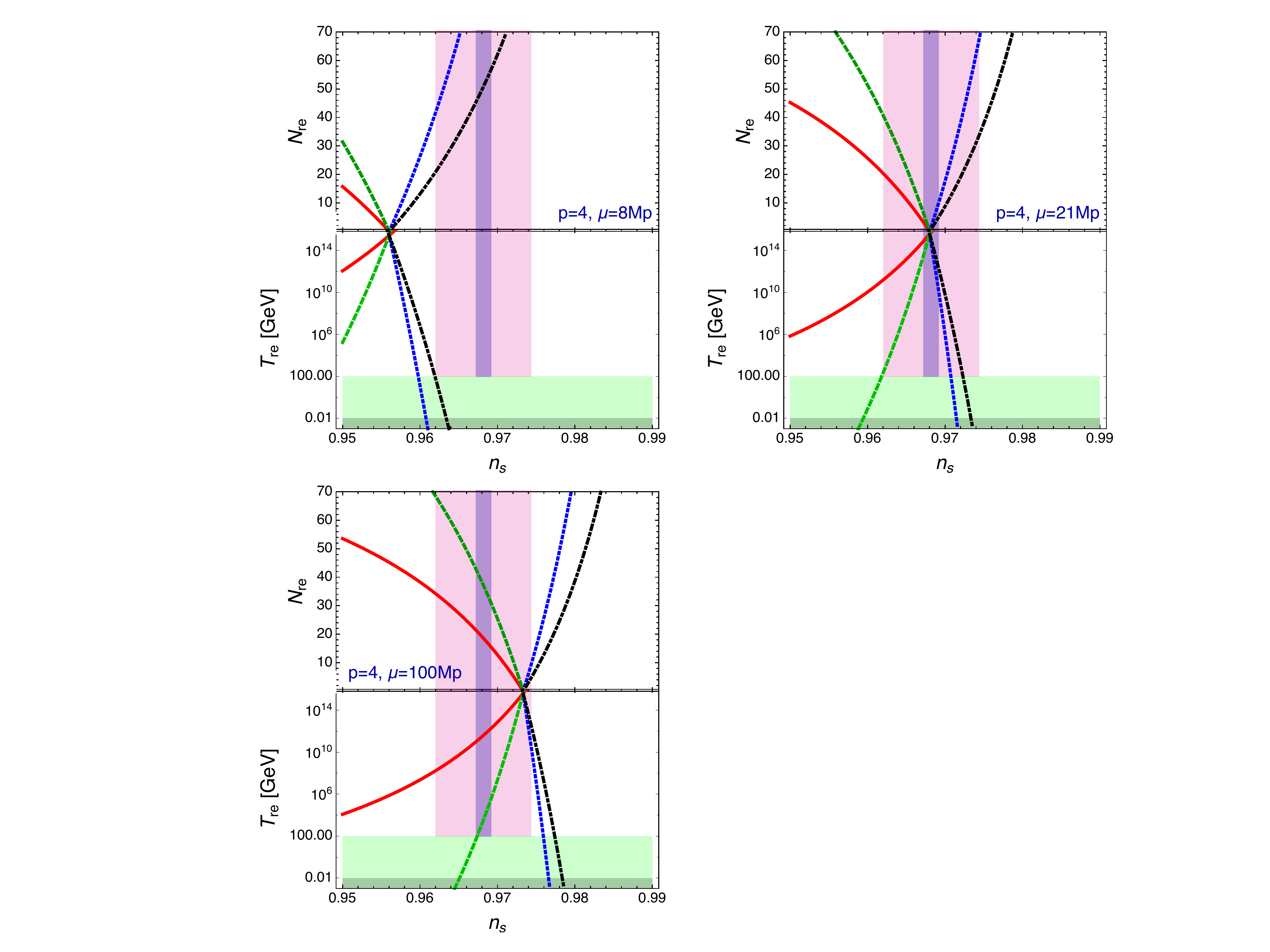}
\caption{Plots of $N_{re}$ and $T_{re}$, for hilltop inflation with $p=4$ and for three different values of $\mu$. Shading is as in Fig.~\ref{fig:a}
}
    \label{fig:i}
\end{figure}

\noindent The plots for hilltop inflation are derived using a numerical procedure and therefore they convey more information than one would obtain with the above analytic results since they cover a range in which the latter would not apply (i.e. for smaller values of $\mu$)\footnote{{Notice that hilltop inflation was previously studied in the context of reheating in \cite{Martin:2010kz,Martin:2014vha,Martin:2014nya}.}}.\\

\noindent Note that for $p = 1$ the potential is just a straight line, and so should give the same predictions as the $V \propto \phi$ inflation model considered above. In Figs.~\ref{fig:g}-\ref{fig:i} we therefore plot $N_{re}$ and $T_{re}$ for various reheating scenarios parametrized by $w_{re}$, and various values of $\mu$ for $p = 2$ and larger.\\

\begin{figure}
\centering
    \includegraphics[width=14cm]{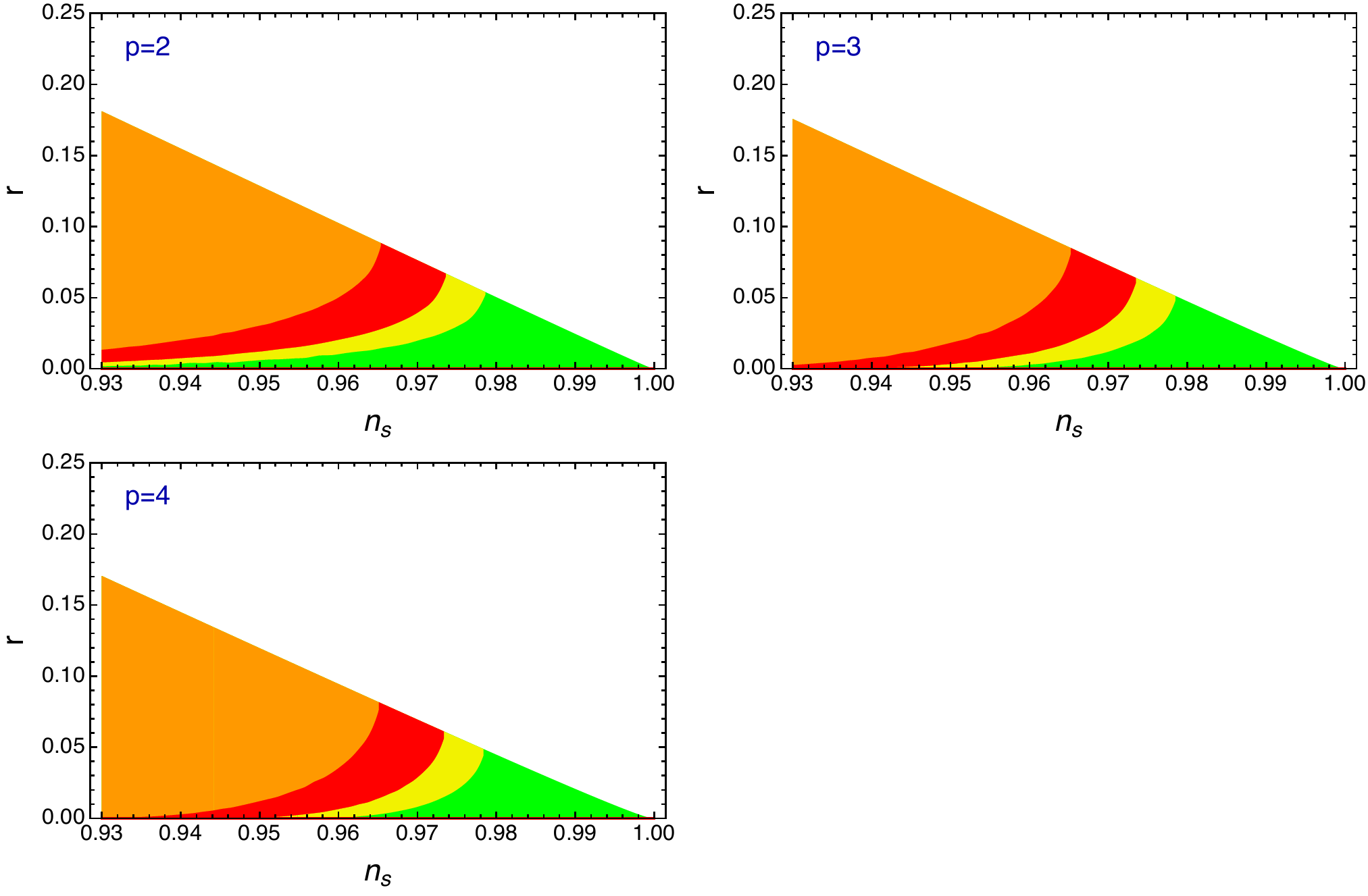}
\caption{Plot of the parameter space in the $r$ vs. $n_{s}$ plane for the three hilltop models with $p =2$, 3, and 4. Shading is as in Fig.~\ref{fig:b} }
    \label{fig:j1}
\end{figure}

\begin{figure}
\centering
    \includegraphics[width=14cm]{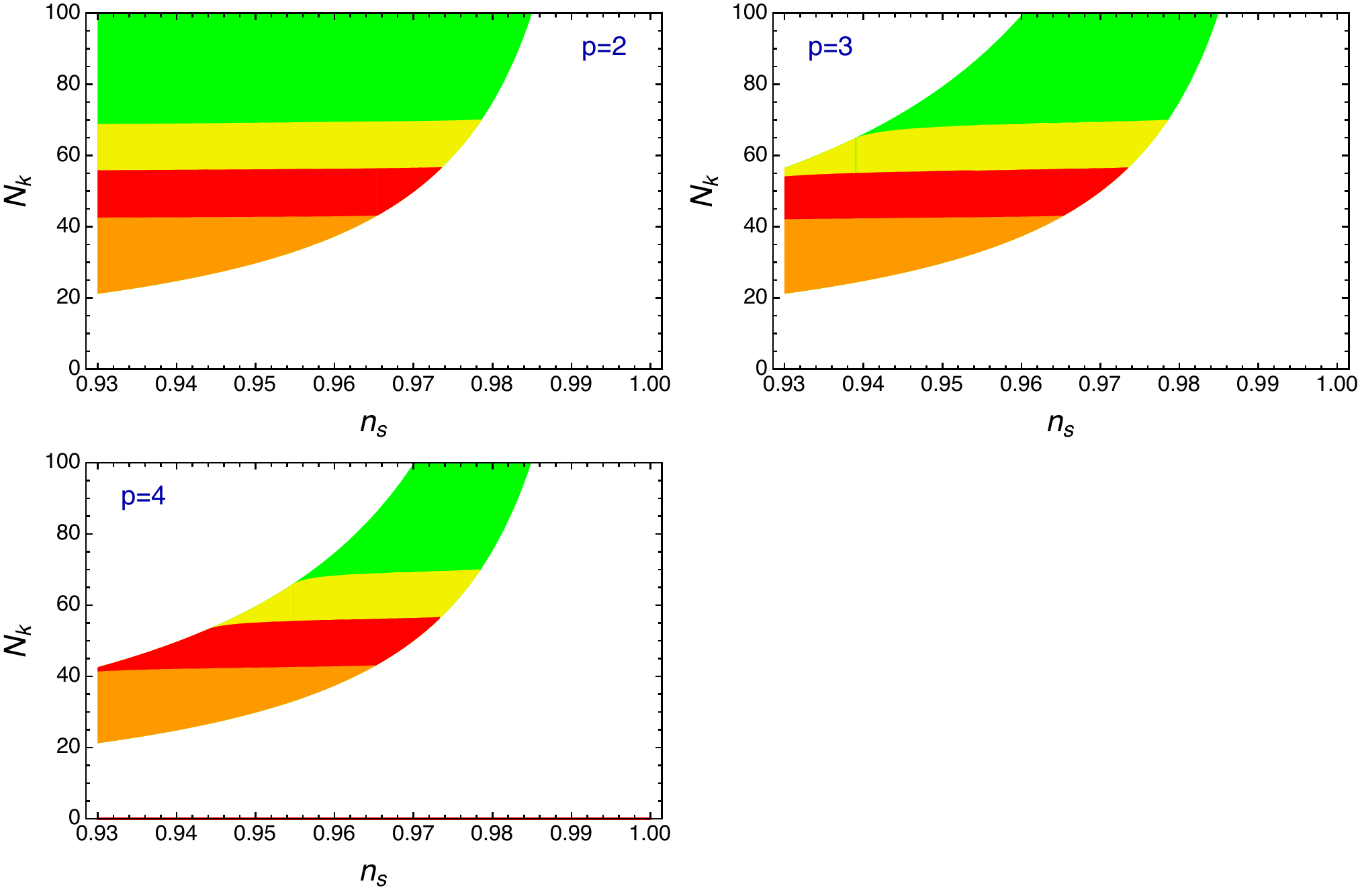}
\caption{Plot of the parameter space in the $N_k$ vs. $n_s$ plane for the three hilltop models with $p =2$, 3, and 4. Shading is as in Fig.~\ref{fig:b} }
    \label{fig:j2}
\end{figure}

\noindent Just as with natural inflation, hilltop inflation has two free parameters, in this case $M$ and $\mu$.  This extra freedom means for each different $p$ value, there are $\mu$ values that are readily consistent with Planck data and $\mu$ values that are not. One can give bounds on $\mu$ for each $p$ model such that reasonable reheating solutions exist, and these results are shown in Figure \ref{fig:j1} and \ref{fig:j2}.  \\

\noindent  As with the bound on $f$ for natural inflation, there is a minimum on $\mu$ required for $p=2$ to get any solution at all, even before reheating is considered, and that bound is a function of $n_s$. This $\mu_{min}$ corresponds to $r \rightarrow 0$, so in this case there is no minimum on $r$ before reheating is considered. For $p=3$, $4$ there is no such minimum on $\mu$ to get a solution in the regime $\mu\geq M_{P}$. There appears to be no observational constraint from Planck for very large values of  $\mu$. \\

The upper bounds in the plots in Figure \ref{fig:j1} (the lower bounds in Figure \ref{fig:j2}) correspond to larger $\mu$, and the lower bounds in Figure \ref{fig:j1} (or the upper bounds in Figure \ref{fig:j2}) correspond to smaller $\mu$. Using the 2$\sigma$ bounds on $n_s$ and requiring $w_{re} \leq 1/3$ gives the following lower bounds on the tensor-to-scalar ratio: $r \geq 0.02$ (for $p=2$) and $r \geq 0.007$ ($p=3$) and $r\geq 0.003$ ($p=4$). Using the central value of $n_s$ and requiring $w_{re} \leq 1/3$ gives the bounds: $r \geq 0.03$ (for $p=2$ and $p=3$) and  $ r \geq 0.02$ ($p=4$). The region in parameter space that is associated with these more likely values of $w_{re}$ then allows for fairly small $r$ values.


\section{Discusion and Conclusions}
\label{sec8}

\indent Inflation includes a wide variety of models that give similar predictions for the fairly small number of available inflationary observables. The physics of reheating can provide an additional opportunity to break this degeneracy. While CMB fluctuations themselves do not supply direct probes of the physics during  the reheating era, the details of reheating affect the predictions for inflation (and vice versa) because they determine the nature of the cosmic thermal history after inflation (see e.g., \cite{Martin:2006rs,Lorenz:2007ze,Martin:2010kz,Adshead:2010mc,Mielczarek:2010ag,Easther:2011yq,Dai:2014jja,Martin:2014nya}). Although we do not know exactly what occurred during reheating, we can make reasonable assumptions such as that the average equation of state during reheating was very likely between 0 and $1/3$. This leads to independent constraints on observables like $n_s$ and $r$, that can then be tested against CMB data.\\
\indent One can parametrize our ignorance about reheating in terms of an equation of state, $w_{re}$, a length $N_{re}$ (measured in terms of number of e-folds elapsed from the end of inflation), and a final temperature, $T_{re}$. For any given inflationary model, one can write relations between the specific model parameters, the amplitude of the scalar power spectrum $A_{s}$, the spectral index $n_{s}$, and the reheating parameters ($w_{re}$, $N_{re}$ and $T_{re}$). These relations are derived by accounting for the total expansion history between the time the observable CMB modes crossed outside the Hubble radius during inflation and the time of observation, and employing a continuity equation for the energy density during the different cosmological epochs. We also assume that $w_{re}$ is constant. For single-field models the derivation is particularly straightforward. The main results are summarized in Eq.~(\ref{mp1}) for $T_{re}$ and $w_{re}$ (and in Eq.~(\ref{eq12}) for $N_{re}$ and $w_{re}$) as a function of inflationary parameters and observables. \\
{We consider a broad range for the equation of state parameter, $-1/3 \leq w_{re} \leq 1$, and the corresponding limits on CMB observables for different inflationary models. }
We notice that a $\phi^2$ potential would favor relatively large values of $r$: a reheating model with $w_{re} \leq 1$ implies $r \geq 0.11$; to allow for a reheating model with $w_{re} \leq 1/3$ which is very probable, requires $r \geq 0.14$. Since it appears BICEP2's signal is dust instead of primordial gravitational waves, it is difficult to reconcile $\phi^2$ inflation with the data. 
 We also consider Starobinsky/Higgs inflation, natural inflation and the hilltop models. For Starobinsky and Higgs inflation, requiring $w_{re}\leq 1/3$ corresponds to $r\geq 0.004$. Because natural and hilltop inflation models have two free parameters, there are ranges of parameter space that can fit well the data for any value of the reheating parameter $w_{re}$. For natural inflation, we find that Planck's 2$\sigma$ bound on $n_{s}$ favors a tensor-to-scalar ratio $r \geq 0.05$ for $w_{re} \leq 1/3$ (Fig.~\ref{fig:f}). For the same range of $w_{re}$, the hilltop model, on the other hand, allows for smaller $r$ values, specifically $r \geq 0.02$ for $p=2$, $r \geq 0.007$ for $p =3$, or $r \geq 0.003$ for $p =4$ (Fig.~\ref{fig:j1}). \\
 
\begin{figure} 
\centering
    \includegraphics[width=11cm]{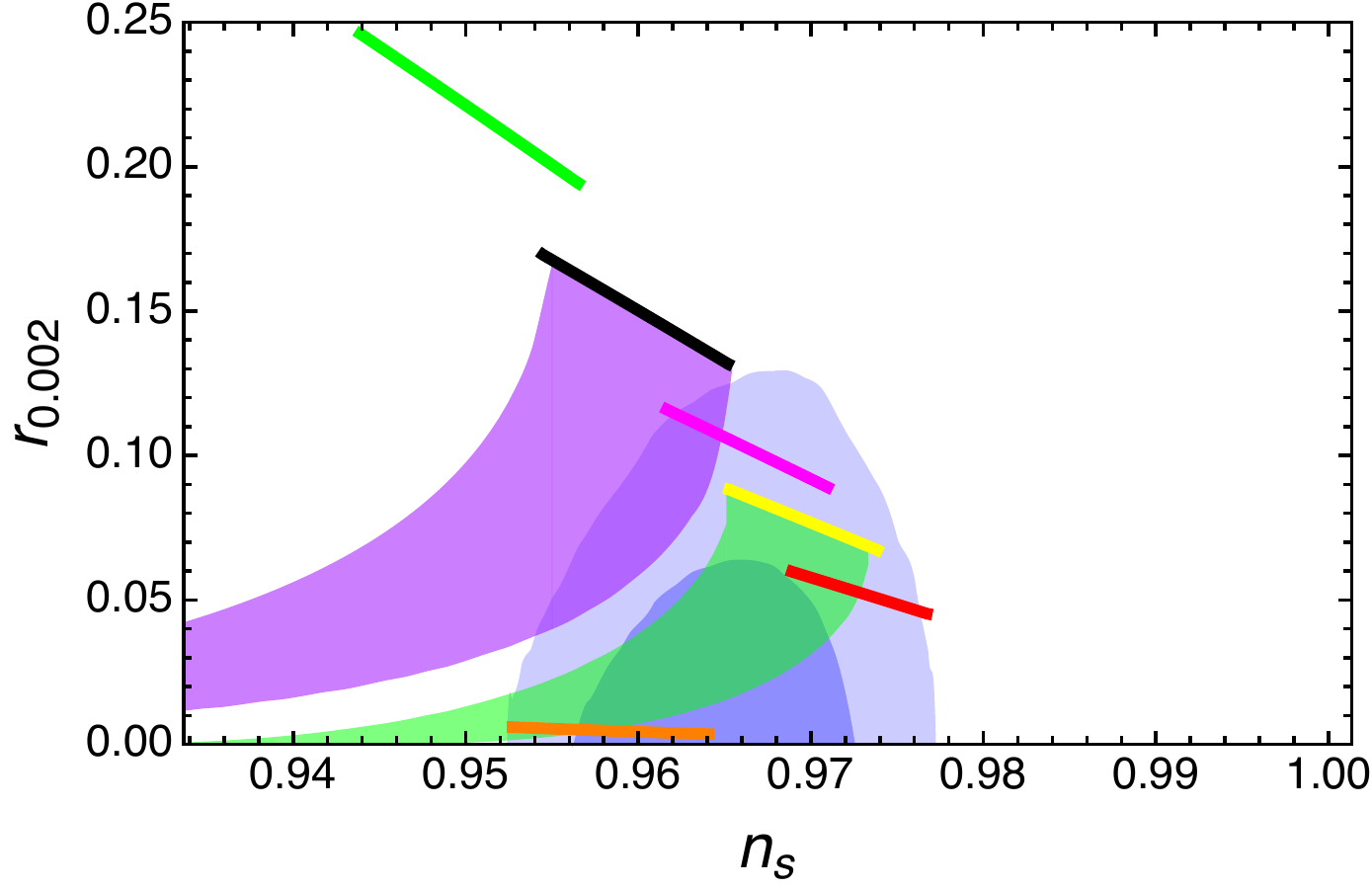}
\caption{We recreate a version of the Planck Figure 12 in their recent Constraints on Inflation paper, using their 1 and 2 $\sigma$ TT, TE, EE + lowP constraints on $n_s$ and $r$ \cite{Ade:2015oja}, but plotting the parameter space for models such that there exists a reheating solution for $0 \leq w_{re} \leq \frac{1}{3}$, as opposed to Planck's choice of parameter space for which there is a solution with $N_k$ between 50 and 60. Following the conventions in Planck's version of the plot, the green line is $\phi^3$, the black is $\phi^2$, the pink is $\phi^{4/3}$, the yellow $\phi$, the red $\phi^{2/3}$, the orange Starobinsky/ Higgs model, the puple region is natural inflation, and the green region is the quartic hilltop model.}
\label{fig:aa}
\end{figure}

We show this parameter space in Fig.~(\ref{fig:aa}), where we recreate a version of the Planck Figure 12 in their recent Constraints on Inflation paper, using their 1 and 2 $\sigma$ TT, TE, EE + lowP constraints on $n_s$ and $r$ \cite{Ade:2015oja}. To get their model parameter space they impose $N_k$ between 50 and 60. Instead here, we don't specify $N_k$ but plot the parameter space for which there exists a reheating solution with $0 \leq w_{re} \leq \frac{1}{3}$. Constraining models in this way, using $w_{re}$, is a nice model dependent but straightforward and well motivated way of representing the parameter space.  \\ 
 
\indent {To conclude we find that considering broad, well-motivated physical constraints on the reheating equation of state indeed allows one to narrow the viable parameter space for inflation models, offering an improvement over merely specifying whether or not an inflation model can reproduce the correct predictions at the pivot scale. These methods will become increasingly effective with future more precise CMB data.}\\

\noindent{{\it �Note added: �\\Our analysis was initially performed considering the Planck 2013 results for the scalar power spectrum parameters. Just after completion, but before submission, Planck released their 2015 data, so we have updated our analysis using the new observational bounds on $n_{s}$ and $A_{s}$. All presented results are now based on the Planck 2015 data. �While completing the first version of this work, it was brought to our attention that a similar approach was carried out by \cite{Munoz:2014eqa}. Some of the results on natural inflation were reproduced in \cite{Munoz:2014eqa} and, where there is overlap, we find agreement if we consider the Planck 2013 bounds on the scalar power spectrum parameters. Furthermore, when this work was near completion, two papers concerning Higgs inflation and reheating were released \cite{Cai:2015soa, Gong:2015qha}. We find agreement with the results of \cite{Cai:2015soa} if we consider their pivot scale ($k_{p}=0.002 Mpc^{-1}$ as opposed to $0.05 Mpc^{-1}$). }}

\acknowledgments

It is a pleasure to thank M. Liguori and J. Wang for useful correspondence. This research was supported in part by a grant from the DOE.

\bibliographystyle{JHEPmodplain}
\bibliography{references}

\end{document}